\documentclass[lettersize,journal]{IEEEtran}
\usepackage{amsmath,amsfonts}
\usepackage{algorithm}
\usepackage{algpseudocode}
\usepackage{array}
\usepackage{comment}
\usepackage[caption=false,font=normalsize,labelfont=sf,textfont=sf]{subfig}
\usepackage{bm}
\usepackage{url}
\usepackage{graphicx}
\usepackage{xcolor}
\usepackage{cite}
\usepackage[hidelinks]{hyperref}
\usepackage{orcidlink}
\usepackage{bbm, dsfont, amssymb}
\newtheorem{theorem}{Theorem}
\newtheorem{lemma}{Lemma}

\newtheorem{prop}{{Proposition}}
\newtheorem{remark}{Remark}
\DeclareMathOperator*{\argmax}{arg\,max}
\DeclareMathOperator*{\argmin}{arg\,min}
\newenvironment{Proof}[1]{\medskip\par\noindent{\bf Proof:\,}\,#1}{{\mbox{\,$\blacksquare$}\par}}
\begin{document}

\title{Strategic Persuasion Through Information Timeliness}
\author{%
Ahmet Bugra Gundogan\,\orcidlink{0009-0002-8878-5633},%~\IEEEmembership{Graduate Student Member,~IEEE,}
\quad Melih Bastopcu\,\orcidlink{0000-0001-5122-0642},~\IEEEmembership{Member,~IEEE}%
\thanks{The authors are with the Department of Electrical and Electronics Engineering, Bilkent University, Ankara, Türkiye (e-mails: bugra.gundogan@bilkent.edu.tr; bastopcu@bilkent.edu.tr).}
       
\thanks{This work was supported in part by the TUBITAK 2232-B program (Project No: 124C533). A part of this paper will be presented at the 23rd IFAC World Congress, Busan, Republic of Korea, August 23-28, 2026, \cite{Gundogan2025arxiv}.}}

\maketitle

\begin{abstract}
We study a dynamic strategic communication problem in which a sender
controls the timing of truthful updates from binary continuous-time Markov
sources. The receiver chooses between a zero-order-hold estimator that
follows the sender's updates and a prior-only default estimator, aiming to
maximize a weighted correct-estimation utility. In contrast, the sender
seeks to persuade the receiver to estimate the state as 1, regardless of
the true state. This misalignment leads to a Stackelberg game in which the
sender, as the leader, commits to state-dependent Poisson update rates,
and the receiver, as the follower, decides whether to follow the sender's
messages. The sender maximizes the long-term average time that the
receiver's estimate equals 1, subject to a conditional intensity budget
and a participation constraint (PC) ensuring that following the sender's
messages does not degrade the receiver's average utility relative to its
prior information. For a single source, we show that the sender's optimal
policy allocates a minimum state-0 update intensity to the undesired state-0, just enough to
satisfy the PC, and the remaining budget to the desired state-1. For
multiple sources with heterogeneous minimum state-0 update intensities, we develop a branch-and-bound algorithm that typically avoids exhaustive
search. Finally, we extend the solution to multiple receivers over
dedicated channels. Our results show that controlling timeliness alone
enables the sender to persuade the receiver and increase its utility.
\vspace{-0.2cm}
\end{abstract}

\begin{IEEEkeywords}

Information design, persuasion through information timeliness, continuous-time Markov chains, Stackelberg games, strategic communication.
\end{IEEEkeywords}
\vspace{-0.2cm}
\section{Introduction}
\vspace{-0.15cm}
\IEEEPARstart{I}{n} public markets, \emph{misrepresentation is illegal}
\cite{sec10b5} while \emph{delaying truthful disclosure can be lawful} under specified conditions \cite{euMAR17}. This distinction creates a setting in which the timing of truthful information disclosure can itself become a strategic instrument. Motivated by this, we model a firm whose fundamentals switch between ``favorable'' (denoted as state-1) and ``unfavorable'' (denoted as state-0) regimes as a binary continuous-time Markov chain (CTMC). The sender (owners or management) cannot lie about outcomes, but can lawfully manage \emph{when} to release verifiable updates by communicating more frequently when fundamentals are good (with rate \(s\)) and more slowly when they are bad (with rate \(c\)), subject to the owner's total conditional-intensity budget \(R\). As another example, we can consider a news provider that supports a campaign but reports only true updates. The campaign's ``momentum'' can be similarly modeled using binary unfavorable ($0$) and favorable ($1$) states. The news provider cannot falsify content, but it can control when truthful reports are aired by releasing favorable news at a higher rate and unfavorable news at a lower rate. These state-dependent reporting rates are chosen subject to a conditional-intensity provisioning budget, which limits the total rate capacity allocated to the two reporting modes. Thus, the provider strategically controls the timing of truthful news rather than its content.  The receivers (investors in the prior example, or the news followers in the latter example) can adopt the firm's (or the news provider's) messages only if doing so does not worsen their long-run
average utility relative to their prior knowledge, which is a participation-type
condition that we call the participation constraint (PC). However, the firm's (or the news agency's) goal is to use the \emph{timeliness} of the information to maximize the fraction of time that the audience believes that the fundamentals (or the campaign, in the latter example) are doing well, subject to the budget \(R\) and the PC. Motivated by these examples, the key research question that we investigate in this work is:

\textit{``By controlling only the timing of the information provided to the receiver, can the information provider (the sender) persuade the receiver to act in a way that the provider's utility is maximized?''}

We study this question within a restricted real-time architecture in which the sender commits to
state-dependent Poisson update intensities and the receiver chooses between a zero-order-hold estimator
based on the sender’s updates and a prior-only default estimator. The resulting equilibrium therefore
characterizes persuasion within these stationary, memoryless policy classes. Our work is related to two distinct strands of strategic information transmission (SIT). Crawford and Sobel~\cite{crawford1982strategic} study strategic communication through cheap talk, in which the sender does not commit in advance to a signaling rule.  In contrast, Bayesian persuasion~\cite{kamenica2011bayesian} considers a sender that commits to an information-revelation policy to influence the receiver's action. Our model follows the latter commitment-based structure: the sender first commits to state-dependent information-revelation rates, and the receiver subsequently chooses its best response. However, unlike classical Bayesian persuasion, where the sender designs what information to reveal, the strategic instrument in our setting is \emph{timeliness}: the sender controls \emph{when} truthful information is released over a dynamically evolving CTMC. The receiver follows messages only when doing so does not worsen its long-run average utility relative to its prior information.

In dynamic approaches, Che et al. \cite{che2023dynamic} study frictions and timing costs, yielding Markov-perfect outcomes and links to static benchmarks. When the sender commits to a disclosure policy over a Markov-evolving
state, greedy policies are optimal for a class of problems
\cite{renault2017optimal}, while \cite{au2015dynamic} studies sequential
disclosure to a privately informed receiver. In more applied settings, dynamic persuasion has been studied with exogenous signals shaping timing incentives \cite{bizzotto2021outside}, receiver search and inspection generating persuasion-acquisition feedback \cite{yao2023dynamicsearch}, partial sender knowledge motivating ``starting rough'' \cite{nuta2024starting}, quadratic state-dependent costs in Gaussian models \cite{sayin2022quadratic}, and continuous-time filtering/control approaches that capture belief dynamics \cite{aid2025filtering}. Related notions of ``timeliness'' arise in models of optimal waiting \cite{orlov2020persuading} and in interim disclosure between mandatory announcement times \cite{gietzmann2023silence}. \emph{Closest to our setting}, Ely \cite{ely2017beeps} shows how a sender \emph{schedules} truthful disclosures about an evolving state to shape the
receiver's actions over time. Ashkenazi-Golan et al.\ \cite{ashkenazi2023markovtwo} study a two-state Markov environment with a myopic receiver and characterize intertemporal disclosure/silence rules. Farhadi and Teneketzis \cite{farhadi2022dynamic} study a principal who
sequentially discloses information about a two-state Markov chain with an
absorbing bad state, so as to delay a strategic detector from detecting
the jump to the bad state. Lehrer and Shaiderman \cite{lehrer2025markov} analyze Markovian persuasion with stochastic revelations.

This rate-based formulation can also model real-time systems in which the timing of information plays a critical role. Recently, age of information (AoI) has been introduced to measure the timeliness of information in communication systems \cite{Kaul2011, Yates12}. The timely remote estimation problem for a Wiener process under a sampling-rate constraint has been considered in the seminal work of \cite{Sun2020}. The timely tracking of Poisson counting processes and of infection status with exponential time intervals has been studied in \cite{Bastopcu_google, bastopcu2022using}. Recently, information sources have been modeled as Markov chains, and remote estimation problems have been studied to minimize the age of incorrect information (AoII) and related semantic metrics in \cite{maatouk2020age, cosandal2024multi, Pappas2025, luo2025roleagesemanticsinformation, saurav2025monitoring}. More specifically, \cite{LuoRemote_estimation} studies the minimization of the age of false and missed alarms in remote estimation of a binary Markov source. In~\cite{ayan2025ageawarecsiacquisitionfinitestate}, the authors study age-aware CSI acquisition over a finite-state Markovian channel, balancing data transmission against the acquisition of fresh CSI. Timely task processing under state-dependent worker performance has been analyzed in the context of task completion efficiency in \cite{sariisik2025maximizeefficiencysystemsexhausted}. Related work studies revenue-maximizing job submission to a queried Markov machine based on the age of its estimated state~\cite{liyanaarachchi2025ageestimatessubmitjobs}.
Unlike timely remote estimation problems in the AoI literature, in our work, the sender and the receiver have misaligned objectives; consequently, by strategically adjusting the timeliness of updates, the sender seeks to persuade the receiver to maintain its estimate in the desired state.

We summarize our main contributions as follows:
\begin{itemize}
    \item We introduce a dynamic persuasion framework in which a sender
    influences a receiver's real-time estimate of binary CTMC sources solely by
    controlling the \emph{timing} of truthful updates. We formulate the interaction as a Stackelberg game and derive a closed-form PC, which reduces to a minimum
    state-0 sampling rate $c_{i,\min}$ for each source (Section~\ref{sect:system}).
    \item In Section~\ref{sect:solution1}, we consider the single source case, where we explicitly characterize the Stackelberg equilibrium: whenever $R\!>\!c_{\min}$, the optimal sender policy allocates exactly the minimum rate $c_{\min}$ to state-$0$ updates to satisfy the PC and devotes the entire remaining budget to state-1 updates. When $R\leq c_{\min}$, we show that the game admits multiple Stackelberg
equilibria, all yielding the sender zero utility.
    \item  In Section~\ref{sect:multisourceproblem}, we extend our analysis to the multi source  setting, where each source may require a different $c_{i,\min}$ to satisfy its PC. We demonstrate that, for any fixed set of sources, the sender's rate-allocation problem for state-1 updates is a convex optimization problem.
    \item In Section~\ref{sec:efficient_algorithm}, we provide a branch-and-bound
algorithm that finds the sender's globally optimal update rate allocation;
its worst-case complexity equals that of exhaustive search, while
dominance-based pruning reduces the search to $\mathcal{O}(n)$ nodes when the
sources admit a total dominance order, and eliminates provably suboptimal
active sets in intermediate cases.
    \item In Section~\ref{sec:multi_receivers_bias}, we extend the problem to a multi source  and multi receiver setting, where each receiver communicates with the sender over a dedicated private channel and may have a heterogeneous utility bias. We then solve the resulting problem optimally using the algorithms developed in the previous section.
    \item Finally, in Section~\ref{Num_result}, we provide illustrative numerical results demonstrating that the sender can achieve persuasion solely by controlling the timeliness of information.
\end{itemize}
\begin{figure}[tb]
  \centering
  \includegraphics[width=0.85\columnwidth]{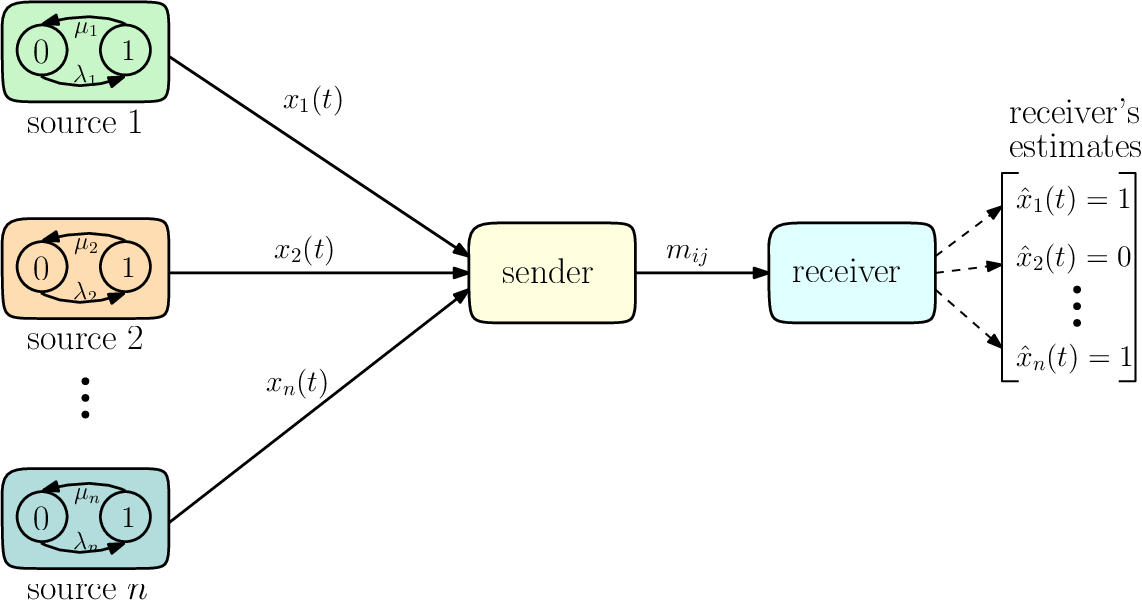}\vspace{-0.25cm}
  \caption{Communication system with $n$ sources, a sender, and a receiver.}
  \label{fig:system}
  \vspace{-0.35cm}
\end{figure}

\vspace{-0.3cm}
\section{\!\!System Model and Problem Formulation}\label{sect:system}
\vspace{-0.1cm}
In this work, we consider a system composed of $n\geq 1$ information sources, a sender, and a receiver. Here, each source denoted  by $I_i$ for $i = 1,\dots,n $  generates binary information streams (0's or 1's) with time-varying dynamics. More specifically, the binary information at source $I_i$ follows a two-state CTMC where the transition from state-0 to state-1 happens with rate $\lambda_i\!>\! 0$ and from state-1 to state-0 with rate $\mu_i\! >\!0$ as shown in Fig.~\ref{fig:system}. We assume that the source processes are mutually independent. Moreover, since $\lambda_i>0$ and $\mu_i>0$, the two-state CTMC associated with each source $i$ is irreducible and therefore admits a unique stationary distribution.

We denote source $i$'s state at time $t$ as $x_i(t)\in\{0,1\}$. Both the sender and the receiver know $(\lambda_i, \mu_i)$ for all $i$, but only the sender is capable of continuously observing these sources and sharing them with the receiver. On the other hand, since the receiver does not observe $x_i(t)$, with its prior knowledge, the receiver will only know the steady-state distribution of the CTMC, which is given by
\begin{align}\nonumber\\[-20pt]
\pi^i_{0} = \frac{\mu_i}{\mu_i + \lambda_i},\qquad
\pi^i_{1} = \frac{\lambda_i}{\mu_i + \lambda_i}.
\label{eq:steady_state_probs}\\[-20pt]\nonumber
\end{align}

The receiver would like to estimate the sources' states as accurately as possible. Based on the prior knowledge of the sources' rates and the information obtained from the sender, the receiver forms a real-time estimate about source $i$'s state, denoted by $\hat{x}_i(t)$, at time $t$. With the sources' states and their corresponding estimates, the receiver will obtain the following utility $ u_i(t)$ from source $i$:
\begin{align*}\\[-19pt]
   u_i(t) \!= \!q  \mathbbm{1} (x_i(t)\!\! =\! 0, \hat{x}_i(t) \!\!=\! 0) \!+\!(1\!-\!q)  \mathbbm{1}(x_i(t)\! =\! 1, \hat{x}_i(t)\! = \!1), \\[-19pt]
\end{align*}
which is shown in Table~\ref{Table:rec_utility} where where $\mathbbm{1}(\cdot)$ denotes the indicator function, which is equal
to $1$ if the statement in its argument holds and to $0$ otherwise.
\begin{table}[t]
\vspace{-0.3cm}
   \caption{The receiver's utility function $u_i(t)$.}
    \label{Table:rec_utility}
    \vspace{-0.1cm}
    \centering
    \begin{tabular}{c|c|c}
        $x_i(t)\backslash \hat{x}_i(t)$ & 0     & 1     \\ \hline
        0                        & $q$   & 0     \\ \hline
        1                        & 0     & $1-q$ \\
    \end{tabular}
    \vspace{-0.35cm}
\end{table}  

In other words, when $x_i(t) = 0$ and $ \hat{x}_i(t) = 0$, the receiver will obtain a weighted reward, that is, $u_i(t) = q$ and when $x_i(t) = 1$ and $ \hat{x}_i(t) = 1$, it will receive $u_i(t) =1- q$ from source $i$ where $0<q<1$. When there is no information provided, the receiver will only know the steady state distribution of $x_i(t)$, i.e., $(\pi^i_{0},\pi^i_{1})$ for all $i$ in (\ref{eq:steady_state_probs}). We assume that $ q\pi^{i}_{0} = \frac{q\mu_i}{\mu_i + \lambda_i} > \frac{(1-q)\lambda_i}{\mu_i + \lambda_i} =(1-q) \pi^i_{1}$ for all $i$. As a result, when there is no information provided, the receiver's default estimation is $\hat{x}_i(t) = 0$ for all $t$. Similarly, the sender's utility obtained from source $i$ at time $t$ is denoted as $v_i(t)$ and is given in Table~\ref{Table:sender_utility}. As opposed to the receiver, the sender's utility will be equal to 1 only when the receiver's estimate is  $\hat{x}_i(t) = 1$ irrespective of  source $i$'s state.  
\begin{table}[t]
   \centering
    \caption{The sender's utility function $v_i(t).$}\label{Table:sender_utility}
\vspace{-0.2cm}
    \begin{tabular}{c|c|c}
        $x_i(t)\backslash \hat{x}_i(t)$ & 0     & 1     \\ \hline
        0                        & 0   & 1     \\ \hline
        1                        & 0     & 1 \\
    \end{tabular}
\vspace{-0.4cm}
\end{table}

Different from most traditional communication literature where the sender and the receiver have aligned goals, here, we note from Tables~\ref{Table:rec_utility} and \ref{Table:sender_utility} that the sender's and the receiver's utility functions are different and the sender's utility also depends on the receiver's estimate at time $t$. More specifically, while the receiver wants to know the states as accurately as possible and thus, maximize a weighted correct-estimation utility, the sender wants the receiver to always estimate the state as $\hat{x}_i(t) =1$ for all $t$.

To model such a system, we consider a setting where the sender shares the sources' states with the receiver at random times. When source $i$'s state is equal to 0, we model the sender’s inter-transmission times as exponentially distributed with rate $c_i\geq 0$. Similarly, when the source $i$'s state is equal to 1, we model the sender’s inter-transmission times as exponentially distributed with rate $s_i\geq 0$. Let us denote the time instant at which the sender sends the $j$th update (where $j\geq1$) about the $i$th source's  state as $t_{i,j}$. By denoting the sender's $j$th message about source $i$'s state as $m_{ij} = x_i(t_{i,j})$, under the condition that the receiver follows the sender's messages (which we will specify precisely), the receiver will form the following estimate $\hat{x}_i(t)$ at time $t$ based on the received messages: 
\begin{align}\nonumber\\[-21pt]\label{est_process}
    \hat{x}_i(t) = m_{ij}, \qquad t_{i,j}\leq t<t_{i,j+1},\\[-21pt] \nonumber
\end{align}
where we assume that the receiver knows the initial values of the sources, i.e., $m_{i0} = x_i(0)$ and $t_{i,0} = 0$ for all $i$. This initialization is adopted only for notational convenience. Under the ergodicity conditions considered below (in particular, for every source with $s_i>0$ and $c_i\geq c_{i,\min}>0$), the joint process $(x_i(t),\hat{x}_i(t))$ admits a unique stationary distribution. Consequently, the long-term average utilities of both the sender and the receiver are independent of the initial source state and the initial estimate.

As the objectives of the sender and the receiver are different, we formulate the interaction between these agents as a Stackelberg game, in which the sender acts as the leader and the receiver as the follower. In this Stackelberg game, the sender commits to a strategy first by choosing $\bm{\beta} = \{\beta_1, \cdots, \beta_n\}$ where $\beta_i = (s_i,c_i)$ for all $i$.\footnote{In this work, we restrict the sender’s policy space to Poisson sampling and characterize the corresponding Stackelberg equilibrium. Focusing on Poisson sampling policies allows for analytical tractability and has been considered in the literature, such as \cite{Bastopcu_google, Akar_CTMC}. } Then, the receiver observes the sender's information-revelation policy and selects its best response. At this point, the receiver has two options: ({\it{i}}) if the sender's messages about source $i$ lead to an estimate no worse than the initial knowledge in terms of maximizing the receiver's long-term average utility, i.e.,  $\lim_{T\rightarrow\infty}\!\!\frac{1}{T}\!\int_{t=0}^{T}\! u_i(t) dt $, the receiver will follow the sender's messages as in (\ref{est_process}). We denote this policy as $\sigma_{\text{sender}}$. ({\it{ii}}) If following the sender's messages leads to a lower average utility compared to the default policy which uses only the prior information, then the receiver will ignore the sender's messages for source $i$ and use the estimate $\hat{x}_i(t) = 0$ for all $t$ which will give the utility of $ q\pi^{i}_{0} = \frac{q\mu_i}{\mu_i + \lambda_i}$ from source $i$. We denote this policy as $\sigma_{\text{default}}$.

\begin{remark}\label{rem:receiver_policy}
We restrict the receiver's strategy space to the two low-complexity
estimators $\sigma_{\text{sender}}$ and $\sigma_{\text{default}}$:
under $\sigma_{\text{sender}}$, the receiver holds the most recently
received message as its estimate as in (\ref{est_process}), which is
the standard zero-order-hold estimator widely adopted in the remote
estimation and AoII literature
\cite{maatouk2020age,bastopcu2022using,LuoRemote_estimation,Akar_CTMC}.
Accordingly, the PC in (\ref{eqn:IC_constraint}) compares
the \emph{long-term average} utilities of the two admissible
estimators. Thus, the PC is a participation-type constraint rather than a
per-history obedience requirement as in classical Bayesian persuasion
\cite{kamenica2011bayesian}. We note that, since the sampling rates are
state-dependent, silence itself carries information: a fully Bayesian
receiver could track its posterior belief between updates and revert
its estimate once the belief drops below its decision
threshold, thereby achieving a utility no smaller than that of
$\sigma_{\text{sender}}$ and weakening the sender's persuasion power.
Together with the restriction of the sender to Poisson sampling
policies (see footnote~1), this defines a Stackelberg game between two
stationary, memoryless policy classes. Characterizing the equilibrium under a belief-tracking
Bayesian receiver is an interesting direction for future work.
\end{remark}

Finally, we represent the sender's utility function obtained from source $i$ as $J_{S,i}(\beta_i,  BR_i(\beta_i))= \lim_{T\rightarrow\infty}\frac{1}{T}\int_{t=0}^{T}v_i(t) dt$ and the sender's total average utility as $J_S(\bm{\beta},  BR(\bm{\beta})) = \sum_{i=1}^n J_{S,i}(\beta_i,  BR_i(\beta_i))$ which depends on the sender's committed policy $\bm{\beta}$ and the receiver's best response to $\bm{\beta}$ given by $BR_i({\beta}_i)\in\{\sigma_{\text{sender}},$ $ \sigma_{\text{default}}\}$. Similarly, based on the sender's policy, if the receiver follows the sender's messages as in (\ref{est_process}), the receiver will obtain the average utility of $J_{R,i}(\beta_i)=\lim_{T\rightarrow\infty}\frac{1}{T}\int_{t=0}^{T}u_i(t) dt $ from source $i$ where the receiver's estimate $\hat{x}_{i}(t)$ is determined in (\ref{est_process}). Thus, we define the Stackelberg equilibrium as
\begin{align}\nonumber\\[-20pt]\label{eqn:Stackelberg}
    J_S(\bm{\beta}^*,  BR(\bm{\beta}^*) ) \geq & J_S(\bm{\beta},  BR(\bm{\beta}) ) \text{ for all $\bm{\beta}$}\\[-24pt]\nonumber
    \end{align}
where 
\begin{align}\nonumber\\[-24pt]\label{eqn:Stackelberg_2}
BR_i(\beta_i)=
\begin{cases}
\sigma_{\text{sender}}, &
J_{R,i}(\beta_i)\geq
\dfrac{q\mu_i}{\mu_i+\lambda_i},\\
\sigma_{\text{default}}, &
J_{R,i}(\beta_i)<
\dfrac{q\mu_i}{\mu_i+\lambda_i}.
\end{cases}\\[-22pt]\nonumber
\end{align}
In other words, the Stackelberg equilibrium within the stated
sender and receiver policy classes in (\ref{eqn:Stackelberg}) and (\ref{eqn:Stackelberg_2}) is achieved when the sender commits to a policy that will maximize its own utility function by  also considering how the receiver will respond to the sender's committed policy. In~(\ref{eqn:Stackelberg_2}), we adopt optimistic leader-favorable tie-breaking: if the two admissible receiver policies yield the same long-term average receiver utility, the receiver selects $\sigma_{\rm sender}$. For the degenerate no-update policy $\beta_i=(0,0)$, we adopt the convention that the receiver uses $\sigma_{\rm default}$, since the sender provides no updates.
\begin{figure}[tb]
  \centering
  \includegraphics[width=0.37\columnwidth]{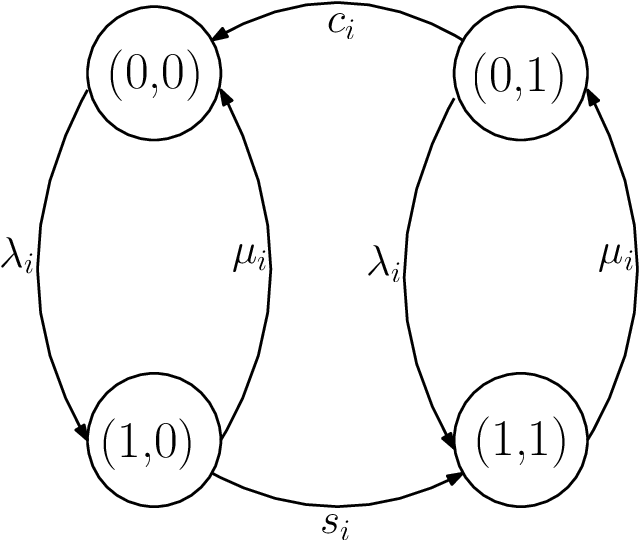}
  \vspace{-0.22cm}
  \caption{Continuous-time Markov chain for $(x_i(t), \hat{x}_i(t))$.}
  \label{fig:CTMC}
  \vspace{-0.4cm}
\end{figure}
  \vspace{-0.5cm}
\subsection{The Receiver's Average Utility Function}
  \vspace{-0.15cm}
As a result of following the sender's messages, source $i$'s state and its estimate at the receiver $(x_i(t), \hat{x}_i(t))$ form a CTMC with four states given as $\{(0,0),(0,1),(1,0), (1,1)\}$ as shown in Fig.~\ref{fig:CTMC}. By dropping source $i$'s index to derive a general expression and assuming that $(s, c) \neq (0, 0)$, similar to the steps in \cite{bastopcu2022using}, we find the unique stationary distribution of the CTMC given by $\pi = \{\pi_{00},\ \pi_{01},\ \pi_{10},\ \pi_{11}\}$. For that, we first write the stationary balance equations as:
\begin{align}\nonumber\\[-21pt] \label{eq:LocalBalanceEq}
   \pi_{00} \lambda &= \pi_{10} \mu + \pi_{01} c, \\[-2pt]
     \pi_{10} \mu + \pi_{10} s &= \pi_{00} \lambda, \\[-2pt]
     \pi_{01} c + \pi_{01} \lambda &= \pi_{11} \mu, \\[-2pt]
     \pi_{11} \mu &= \pi_{10} s + \pi_{01} \lambda .\\[-21pt] \nonumber
\end{align}
Using the above equations and $ 
\sum_{m=0}^{1} \sum_{n=0}^{1}\! \pi_{mn} = 1, 
$ we find the steady-state distribution of the CTMC as:
\begin{align*}\\[-20pt]
\pi_{00} \!= \!\frac{\mu c (\mu\!+\!s)}{\kappa},~ \pi_{01}\! =\! \frac{\mu  \lambda s}{\kappa},~ \pi_{10} \!= \!\frac{\mu  \lambda c}{\kappa},~ \pi_{11}\! =\! \frac{\lambda s (\lambda \!+\!c) }{\kappa},\\[-20pt]
\end{align*}
where $\kappa = (\mu + \lambda)(\mu c + \lambda s + c s) $. Then, from Table~\ref{Table:rec_utility}, we can find the receiver’s
long-term average utility obtained from source $i$ as
\begin{align} \nonumber\\[-20pt]
J_{R,i}(\beta_i)
&= q\,\pi^{i}_{00} + (1-q)\,\pi^{i}_{11} \notag\\[-2pt]
&= \frac{ q\,\mu_i c_i (\mu_i+s_i) + (1-q)\,\lambda_i s_i (\lambda_i + c_i)}
         { (\mu_i+\lambda_i)\bigl(\mu_i c_i + \lambda_i s_i + c_i s_i\bigr)}.
\label{eqn:rec_cost_closed_form}\\[-20pt] \nonumber
\end{align}

Note that
if the messages $m_{ij}$ help the receiver to form an estimate no worse
than the initial knowledge (which will be the compliance/participation condition for the two
admissible receiver policies in our Stackelberg game), the receiver will follow the messages $m_{ij}$ as in (\ref{est_process}) and as a result obtain $\!J_{R,i}(\beta_i)\!$ in (\ref{eqn:rec_cost_closed_form}). Otherwise, the receiver will use $\hat{x}_i(t)\!=\!0$ for all $t$ and obtain the utility of $\frac{q\mu_i}{\mu_i +\lambda_i}$ from source $i$.  
\vspace{-0.45cm}
\subsection{Sender's Persuasion Problem}
\vspace{-0.2cm} 
As noted earlier, the sender's utility function
$v_i(t)$ provided in Table~\ref{Table:sender_utility} depends on the receiver's estimation. As a result, the sender wants to influence the receiver to follow its messages and affect $\hat{x}_i(t)$ in a way to maximize its own utility. To do that, the sender should commit to a policy  $\beta_i = (s_i,c_i)$ such that the receiver's utility $J_{R,i}(\beta_i)$ is greater than or equal to $ \frac{q\mu_i}{\mu_i + \lambda_i}$ which is the PC. When we substitute $J_{R,i}(\beta_i)$ in (\ref{eqn:rec_cost_closed_form}) into $J_{R,i}(\beta_i)\!\geq \!\frac{q\mu_i}{\mu_i + \lambda_i}$ and perform some algebraic manipulations, under the assumption that $s_i\!>\!0$, we obtain
\begin{align}\nonumber\\[-20pt]\label{eqn:IC_constraint}
    c_i\geq c_{i,\min} =  \frac{q\mu_i}{1-q}-\lambda_i. \\[-20pt]\nonumber
\end{align}
Due to our assumption that the receiver's initial estimation without the sender's information is equal to 0, that is $ q\pi^{i}_{0} = \frac{q\mu_i}{\mu_i + \lambda_i} > \frac{(1-q)\lambda_i}{\mu_i + \lambda_i} =(1-q) \pi^i_{1}$, $c_{i,\min}$ will always be positive, i.e., $c_{i,\min}>0$. On the other hand, when $s_i = 0$, this constraint is automatically satisfied for all $c_i$. We impose an additive budget on the state-conditioned update-intensity parameters, $ \sum_{i=1}^{n} (c_i + s_i) \leq R.$\footnote{Thus, $R$ prices the two conditional intensity controls rather than the realized stationary number of transmissions. Under policy $(s_i, c_i)$, the realized average update intensity of source $i$ is $\pi_{0}^i
c_i +\pi_{1}^is_i$.}

Then, the sender's persuasion problem becomes: 
\begin{align}\nonumber\\[-20pt]
\label{eqn:senders_problem}
\max_{\{ s_i, c_i \}}  \quad & \sum_{i=1}^{n}(\pi^i_{01} + \pi^i_{11})  \!\!= \!\!\sum_{i=1}^{n}\frac{\lambda_i s_i (c_i + \lambda_i + \mu_i)}{(\mu_i + \lambda_i)(\mu_i c_i + \lambda_i s_i + c_i s_i)} \nonumber \\[-2pt]
\mbox{s.t.} \quad & \sum_{i=1}^{n} (c_i + s_i) \leq R \nonumber \\[-2pt]
\quad &  c_i\geq \mathbbm{1}(s_i>0) c_{i,\min}, ~ s_i \geq 0, ~~ i \in\{1,\ldots,n\}.\!\!\!\\[-20pt]\nonumber
\end{align} 
We note from (\ref{eqn:senders_problem}) that the sender has the  conditional intensity budget $R$ such that $\sum_{i=1}^{n} c_i + s_i \leq R$. The second constraint ($c_i\!\geq\! \mathbbm{1}(s_i\!>\!0) c_{i,\min}$) in  (\ref{eqn:senders_problem}) is the PC for each source, and the third constraint is the feasibility constraint. \footnote{In the Stackelberg game formulated in (\ref{eqn:Stackelberg})
and (\ref{eqn:Stackelberg_2}), the sender may also choose $s_i>0$ with
$c_i<c_{i,\min}$ for some source $i$. In that case, the receiver's best
response is $\sigma_{\text{default}}$, and the sender obtains zero
utility from source $i$ while consuming the positive rate $s_i+c_i$.
Such a policy is weakly dominated by allocating $(s_i,c_i)=(0,0)$ to
source $i$, which yields the same zero utility while freeing the rate
$s_i+c_i$; hence, the constraint $c_i\geq\mathbbm{1}(s_i>0)\,c_{i,\min}$
in (\ref{eqn:senders_problem}) entails no loss of optimality. Similarly,
the sources from which the sender seeks no utility are represented by
$(s_i,c_i)=(0,0)$, in which case the corresponding summand in
(\ref{eqn:senders_problem}) is defined to be zero, consistent with the
zero utility obtained under $\sigma_{\text{default}}$. Consequently, the
optimal values of (\ref{eqn:senders_problem}) and the Stackelberg game
coincide.}

In the next section, we provide the Stackelberg equilibrium of the game formulated in (\ref{eqn:Stackelberg}) and (\ref{eqn:Stackelberg_2}) for a single source. 

\vspace{-0.15cm}
\section{Optimal Information Revelation: Single Source and Single Receiver}\label{sect:solution1}
\vspace{-0.15cm}

In this section, we provide an explicit solution to the sender's information-revelation problem in (\ref{eqn:senders_problem}) and thereby characterize a Stackelberg equilibrium of the game formulated in (\ref{eqn:Stackelberg}) and (\ref{eqn:Stackelberg_2}) for a single source.  For convenience, by dropping the source index $i$, we rewrite the sender's optimization problem for a single source, i.e., when $n=1$, as
\begin{align}\nonumber\\[-20pt]
\label{eqn:senders_problem_n_1}
\max_{\{ s, c \}}  \quad & J_S(s,c) =\pi_{01} + \pi_{11}  = \frac{\lambda s (c + \lambda + \mu)}{(\mu + \lambda)(\mu c + \lambda s + c s)} \nonumber \\[-2pt]
\mbox{s.t.} \quad & c + s \leq R \nonumber \\[-2pt]
\quad &  c\geq \mathbbm{1}(s>0) c_{\min}, \quad s \geq 0.\\[-20pt]\nonumber
\end{align}

As seen in (\ref{eqn:senders_problem_n_1}), there is a minimum sampling rate for $c$, denoted as $c_{\min} = \frac{q \mu}{1-q}-\lambda$, arising from the PC, which is strictly positive as mentioned before. In order to find the sender's optimal solution, in the next lemma, we characterize how the sender's utility function behaves with respect to the sampling rates $s$ and $c$.
\begin{lemma}\label{Lemma_1}
 Under the assumption that the PC holds, the sender's utility $J_S(s,c)$ in (\ref{eqn:senders_problem_n_1}) is strictly increasing in $s$ when $c>0$, and strictly decreasing in $c$ when $s>0$.   
\end{lemma} 
\begin{Proof}
We begin the proof by showing that $J_S(s,c)$ is an increasing function of $s$ when $c>0$. For that, the partial derivative of $J_S(s,c)$ with respect to $s$ is given by 
\begin{align*}\\[-20pt]
    \frac{\partial J_S(s,c)}{\partial s} = \frac{\lambda\mu c( \lambda + \mu+c)}{(\mu + \lambda)(\mu c + \lambda s + c s)^2}.\\[-20pt]
\end{align*}
Thus, we have $\frac{\partial J_S(s,c)}{\partial s}\!>\!0$ when $c\!>\!0$, which is indeed the case by the PC, as it implies that $c\geq c_{\min}>0$. Similarly, when $s>0$, $J_S(s,c)$ is decreasing in $c$ since
\begin{align*}\\[-20pt]
    \frac{\partial J_S(s,c)}{\partial c} = -\frac{\lambda \mu s(\lambda+\mu+s)}{(\mu+\lambda)(\mu c+\lambda s + cs)^2}.\\[-20pt]
\end{align*}
As a result, we have $\frac{\partial J_S(s,c)}{\partial c} <0$ when $s>0$, which completes the proof. 
\end{Proof}

Thus, whenever the PC can be met, to maximize its own utility the sender should allocate most of its sampling rate to $s$ and just enough of its sampling rate to $c$ to meet the PC. In the next theorem, we characterize the Stackelberg equilibrium of the game for the single source case.
\begin{theorem}\label{Thm_1}
The single source game admits the following Stackelberg equilibrium characterization:
\begin{itemize}
    \item[($i$)] If $R \leq c_{\min}$, the sender's optimal utility is zero.
    One such Stackelberg equilibrium attaining this utility is 
    $(\beta^*, BR(\beta^*)) = ((0,0), \sigma_{\text{default}})$.
    \item[($ii$)] If $R > c_{\min}$, the sender's optimal policy is
    $\beta^* = (\bar{s}, c_{\min})$ with $\bar{s} = R - c_{\min} > 0$, to
    which the receiver's best response is
    $BR(\beta^*) = \sigma_{\text{sender}}$ under the tie-breaking convention
    in (\ref{eqn:Stackelberg_2}). At this equilibrium, the sender obtains
    the utility
    \begin{align*}
    J_S(\beta^*, BR(\beta^*)) =
    \frac{\lambda \bar{s}\,(c_{\min}+\lambda+\mu)}
         {(\mu+\lambda)\bigl(\mu c_{\min}+ \lambda \bar{s} + c_{\min} \bar{s}\bigr)}.
    \end{align*}
\end{itemize}
In both cases, the receiver's equilibrium utility equals its prior-only
utility, $J_R(\beta^*, BR(\beta^*)) = \frac{q\mu}{\mu+\lambda}$.
\end{theorem}
\begin{Proof}
We begin our proof by considering the case when $R\leq c_{\min}$ and show
that every feasible policy yields the sender zero utility. Consider any
policy $\beta=(s,c)$ with $c+s\leq R$. If $s>0$, then
$c\leq R-s<R\leq c_{\min}$, so the sender cannot meet the PC in
(\ref{eqn:IC_constraint}). Consequently, the receiver would not follow the
sender's messages, i.e., $BR(\beta)=\sigma_{\text{default}}$, and would keep
its estimate $\hat{x}(t)=0$ for all $t$, in which case the sender would
obtain zero utility. If instead $s=0$, the sender never reports state-1, and
thus obtains zero utility under any best response of the receiver, since
$\hat{x}(t)=1$ can hold at most during an initial transient; in this case,
the receiver obtains its default utility under either response. In
particular, when $R=c_{\min}$, the only policies satisfying the PC are those with $s=0$ and $c=c_{\min}$, which fall into the
second case. Since all feasible policies yield the sender the same (zero)
utility when $R\leq c_{\min}$, we choose
$(\beta^*, BR(\beta^*))=((0,0),\sigma_{\text{default}})$ as one such
Stackelberg equilibrium, at which the sender obtains
$J_S(\beta^*, BR(\beta^*))=0$ and the receiver obtains
$J_R((0,0),\sigma_{\text{default}})=\frac{q\mu}{\mu+\lambda}$, which
establishes part~($i$).

Next, we consider the case when $R\!>\! c_{\min}$. In this case, the sender
has a sufficient conditional intensity budget to persuade the receiver to
follow its messages while allocating $\bar{s}=R-c_{\min}>0$ to state-1
updates. By Lemma~\ref{Lemma_1}, the sender's utility is an increasing
function of $s$ and a decreasing function of $c$ when $s>0$ and $c>0$. As a
result, the sender should set $c\!=\! c_{\min}$ to satisfy the PC
and then allocate the remaining sampling rate to $s$ to maximize its own
utility. Hence, when $R> c_{\min}$, the Stackelberg equilibrium is achieved
at $(\beta^*, BR(\beta^*))=((R-c_{\min},c_{\min}), \sigma_{\text{sender}})$,
where $BR(\beta^*)=\sigma_{\text{sender}}$ follows from the tie-breaking
convention embedded in (\ref{eqn:Stackelberg_2}). The corresponding
utilities of the sender and the receiver are given by
$J_S(\beta^*, BR(\beta^*)) = \frac{\lambda \bar{s} (c_{\min}+\lambda+\mu)}{(\mu+\lambda)(\mu c_{\min} + \lambda \bar{s} + c_{\min} \bar{s})}$
with $\bar{s} = R-c_{\min}$ and
$J_R(\beta^*, BR(\beta^*)) =\frac{q\mu}{\mu + \lambda}$, respectively, which
establishes part~($ii$).
\end{Proof}

When $R > c_{\min}$, as a result of applying the sender's optimal policy found in Theorem~\ref{Thm_1}, the receiver follows the sender's messages, i.e., $\sigma_{\text{sender}}$, and the receiver's utility is the same as in the case with only prior information due to the optimistic leader-favorable tie-breaking assumption. However, the sender benefits from applying this policy, as the receiver's estimate equals $1$ for some portion of the time. Building on these insights, we next generalize the results from the single source case to the multi source  setting.

\vspace{-0.35cm}
\section{Optimal Information Revelation: Multi Source and Single Receiver}\label{sect:multisourceproblem}
\vspace{-0.15cm}
In this section, we extend our analysis to a setting where the sender reveals information about multiple sources to the receiver. Our goal is to solve the general persuasion problem with $n\geq 1$ sources in (\ref{eqn:senders_problem}). The sender's information-revelation policy is more involved since the sender should decide which information source it should sample and at which rates. In order to characterize the sender's optimal information-revelation policy, we start with the setting in which the sender's total budget is limited by $R\leq \min_{i\in\{1,\dots,n\}} c_{i,\min}$. 

\begin{lemma}\label{Lemma2}
When $R \leq \min_{i\in\{1,\dots,n\}} c_{i,\min}$, the sender's optimal
utility is zero. One such Stackelberg equilibrium achieving this utility is $\beta_i^* = (0,0)$ and
$BR_i(\beta_i^*) = \sigma_{\text{default}}$ for all $i$, at which the
receiver obtains $\sum_{i=1}^{n}\frac{q\mu_i}{\mu_i+\lambda_i}$.
\end{lemma}
\begin{Proof}
For any feasible policy, every source $i$ with $s_i>0$ satisfies
$c_i \leq R - s_i < c_{i,\min}$, so its PC fails. As a result, the receiver chooses
$BR_i(\beta_i)=\sigma_{\text{default}}$ and the sender obtains $J_{S,i}(\beta_i, BR_i(\beta_i))=0$. On the other hand, every source with
$s_i=0$ yields $J_{S,i}(\beta_i, BR_i(\beta_i))=0$ under any best response of the receiver, by the
argument in the proof of Theorem~\ref{Thm_1}($i$). Hence the sender's
utility at equilibrium is zero, and it can be attained by $\beta_i^*\!=\!(0,0)$ with
$BR_i(\beta_i^*)\!=\!\sigma_{\text{default}}$ for all $i$.  
\end{Proof}

In the remaining part of this section, we focus our attention on the setting where $\!R\!>\!\! \min_{i\in\{1,\dots,n\}} \!c_{i,\min}$. Thus, at least by allocating all of its sampling rate, the sender is capable of persuading the receiver to follow its messages for some sources. To characterize the sender's optimal information-revelation policy, next, we state that the sender allocates a positive sampling rate $s_i\!>\!0$ to source $i$'s state-1 if and only if the sampling rate for source $i$'s state-0 satisfies $c_i = c_{i,\min}$.   
\begin{lemma}\label{Lemma3}
When $R>\min_{i\in\{1,\ldots,n\}} c_{i,\min}$, any optimal policy with
positive sender utility satisfies the following: for every source $i$,
exactly one of a) $(s_i^*,c_i^*)=(0,0)$; or b) $s_i^*>0$ and
$c_i^*=c_{i,\min}$ holds.
\end{lemma}
\begin{Proof}
Since $R>\min_{i} c_{i,\min}$, the policy that assigns
$c_m=c_{m,\min}$ and $s_m=R-c_{m,\min}>0$ to a source
$m\in\argmin_i c_{i,\min}$ and $(0,0)$ to all other sources is feasible and, by Lemma~\ref{Lemma_1}, yields a strictly positive utility. Hence, the sender's optimal policy satisfies $J_S(\bm{\beta}^*, BR(\bm{\beta}^*))>0$, and there exists a source $j$ with $J_{S,j}(\beta_j^*, BR_j(\beta_j^*))>0$, which requires $s_j^*>0$ and $c_j^*\geq c_{j,\min}$. Moreover, by (\ref{eqn:IC_constraint}), the PC threshold $c_{j,\min}$ does not depend on $s_j$; thus, increasing $s_j$ alone never violates source $j$'s PC.

First, suppose that $s_i^*>0$ and $c_i^*<c_{i,\min}$ for some $i$. Then, the PC fails for source $i$, so $J_{S,i}(\beta_i^*, BR_i(\beta_i^*))=0$ while source $i$ consumes the rate $\delta=s_i^*+c_i^*>0$. Note that $i \neq j$ since $J_{S,j}(\beta_j^*, BR_j(\beta_j^*))>0$. By reallocating this rate, i.e., setting $(s_i,c_i)=(0,0)$ and $s_j=s_j^*+\delta$, the total rate and all other sources are unchanged, source $j$'s PC still holds, and $J_{S,j}$ strictly increases by Lemma~\ref{Lemma_1} since $c_j^*\geq c_{j,\min}>0$, contradicting optimality.

Second, suppose that $s_i^*\!>\!0$ and $c_i^*\!>\!c_{i,\min}$ for some $i$. By setting $c_i=c_{i,\min}$ and $s_i=s_i^*\!+\!(c_i^*\!-\!c_{i,\min})$, the total rate is unchanged and the PC for source $i$ still holds with equality. Since $J_{S,i}$ is strictly decreasing in $c_i$ and strictly
increasing in $s_i$ by Lemma~\ref{Lemma_1}, this modification strictly increases $J_{S,i}$, again contradicting optimality. Therefore, if $s_i^*\!>\!0$, then $c_i^*\!=\!c_{i,\min}$.

It remains to consider the case $s_i^*=0$ and $c_i^*>0$ for some source $i$. Since $s_i^*=0$, source \(i\) yields zero sender utility while consuming the positive budget \(c_i^*>0\). Since the optimal sender utility is strictly positive, there exists at least one source \(j\) with \(s_j^*>0\) and, from the preceding arguments, \(c_j^*=c_{j,\min}>0\). We can therefore set \((s_i,c_i)=(0,0)\) and increase \(s_j\) to \(s_j^*+c_i^*\), while keeping all other variables unchanged. This modification preserves the total budget and source \(j\)'s PC, since \(c_j^*=c_{j,\min}\) remains unchanged. By Lemma~\ref{Lemma_1}, increasing \(s_j\) strictly increases the sender's utility, contradicting the optimality of the original policy. Hence, $s_i^*=0$ implies $c_i^*=0$. Combining the above cases, every source $i$ at an optimal policy satisfies either \((s_i^*,c_i^*)=(0,0)\), or \(s_i^*>0\) and \(c_i^*=c_{i,\min}\), which completes the proof.
\end{Proof}

Thus, Lemma~\ref{Lemma3} shows that, at an optimal policy with positive sender utility, each source is either not sampled, or sampled with $s_i>0$ and $c_i=c_{i,\min}$. Accordingly, we introduce a binary activation variable $z_i\in\{0,1\}$ for
each source: $z_i=1$ indicates that source $i$ is active, i.e., sampled with
$s_i>0$ and $c_i=c_{i,\min}$, whereas $z_i=0$ indicates that it is inactive,
i.e., not sampled, with $(s_i,c_i)=(0,0)$. The sender's problem in (\ref{eqn:senders_problem}) can then be reformulated as
\begin{align}\nonumber\\[-23pt]
\label{eqn:senders_problem_mod}
\max_{\{z_i, s_i\}}  \quad & \hat{J}_S(\mathbf{z},\mathbf{s}) = \sum_{i=1}^{n}  z_i\frac{\lambda_i}{\lambda_i + \mu_i}  \frac{s_i (c_{i, \min} + \mu_i + \lambda_i)}{c_{i, \min} \mu_i + s_i \lambda_i + c_{i, \min} s_i} \nonumber \\[-6pt]
\mbox{s.t.} \quad & 
\sum_{i=1}^{n} \left( z_i c_{i, \min} + s_i \right) \leq R\nonumber \\[-4pt]
\quad &  z_i \in \{0, 1\}, ~~~s_i \geq 0,~~~ i \in\{1,\ldots,n\}.\\[-21pt]\nonumber
\end{align}
Thus, $z_i$ captures whether source $i$ is activated, while the corresponding state-$0$ update rate of an active source is fixed at the minimum value required by the PC. Therefore, multiplying the objective function by $z_i$ in (\ref{eqn:senders_problem_mod}) appropriately captures this behavior. Similarly, the  conditional intensity budget constraint can be written as $\sum_{i=1}^{n} \!\!\left( z_i c_{i, \min} \!+\! s_i \right) \!\leq\! R$.

To solve the optimization in (\ref{eqn:senders_problem_mod}), we fix the values of $z_i$ and then solve the resulting optimization problem over the variables $s_i$. With this goal, for a given set of $z_i$'s, we first analyze the convexity of the sender's persuasion problem in (\ref{eqn:senders_problem_mod}) with respect to $s_i$.

\begin{lemma}\label{Lemma4}
For a given set of $z_i$'s,  the sender's information-revelation problem in (\ref{eqn:senders_problem_mod}) is a
convex optimization problem.
\end{lemma}
\begin{Proof}
The first and second derivatives of the sender's utility function are given by 
\begin{align*}
    \frac{\partial \hat{J}_S(\mathbf{z},\mathbf{s})}{\partial s_i}&= z_i \frac{\lambda_ic_{i, \min} \mu_i}{\lambda_i +\mu_i} \frac{(c_{i, \min} + \mu_i + \lambda_i)}{(c_{i, \min} \mu_i + s_i\lambda_i + s_ic_{i, \min} )^2}.\\
    \frac{\partial^2 \hat{J}_S(\mathbf{z},\mathbf{s})}{\partial s_i^2}\!&=\! -2z_i \frac{\lambda_ic_{i, \min} \mu_i}{\lambda_i +\mu_i} \frac{(\lambda_i \!+\! c_{i, \min})(c_{i, \min} + \mu_i + \lambda_i)}{(c_{i, \min} \mu_i + s_i\lambda_i + s_ic_{i, \min} )^3}.
\end{align*}
Since the first derivative is non-negative and the second derivative is non-positive, that is, $\frac{\partial \hat{J}_S(\mathbf{z},\mathbf{s})}{\partial s_i}\geq 0$ and  $\frac{\partial^2 \hat{J}_S(\mathbf{z},\mathbf{s})}{\partial s_i^2}\leq 0$, respectively, we can conclude that the sender's utility is a concave non-decreasing function of $s_i$. Since the conditional intensity budget constraint, $\sum_{i=1}^{n} \left( z_i c_{i,\min} + s_i \right) \leq R$,\footnote{For some given sets of $z_i$'s, we may have $R- \sum_{i=1}^{n}z_i c_{i,\min} < 0$, in which case the problem is infeasible.} and the feasibility constraint, $s_i \geq 0$, define a convex feasible region, the optimization problem in (\ref{eqn:senders_problem_mod}) is convex for any fixed set of $z_i$ values.    
\end{Proof}
For a given set of $z_i$'s, let us denote $\mathcal{S}$ as the set of {\it active} source indices such that $z_i = 1$. Then, the complement $\mathcal{S}^c$ is the set of {\it inactive} source indices with $z_i = 0$. For sources in $\mathcal{S}^c$, we have $s_i = 0$ due to Lemma~\ref{Lemma3}. For the remaining sources in $\mathcal{S}$, we introduce the Lagrangian function \cite{boyd2004convex} for  (\ref{eqn:senders_problem_mod}) to find their optimum update rates $s_i$:
\begin{align*}\\[-21pt]
    \mathcal{L} =& -\sum_{i\in \mathcal{S}}  \frac{\lambda_i}{\lambda_i + \mu_i}  \frac{s_i (c_{i, \min} + \mu_i + \lambda_i)}{c_{i, \min} \mu_i + s_i \lambda_i + c_{i,\min} s_i}\nonumber\\[-3pt]& +\theta \left(\sum_{i\in \mathcal{S}} \left( c_{i, \min} + s_i \right)- R\right) \!- \!\sum_{i\in \mathcal{S}} \nu_i s_i,\\[-21pt]
\end{align*}
where $\theta\geq 0$ and $\nu_i \geq 0$ for all $i$. Next, the KKT conditions are given by 
\begin{align}\nonumber\\[-20pt]\label{eqn:KKT_cond}
\!\!\!\!\frac{\partial \mathcal{L}}{\partial s_i} \!\!=\!\! - \frac{\lambda_ic_{i, \min} \mu_i}{\lambda_i +\mu_i} \frac{(c_{i, \min} + \mu_i + \lambda_i)}{(c_{i, \min} \mu_i \!+ \!s_i\lambda_i \!+\! s_ic_{i, \min} )^2} \!+\! \theta\!-\!\nu_i  \!\!=\!\! 0,\!\!   \\[-20pt]\nonumber
\end{align}
for all $i\in \mathcal{S}$. Then, the complementary slackness (C.S.) conditions can be stated as follows:  
\begin{align}\nonumber\\[-20pt]
    \theta \left(\sum_{i\in \mathcal{S}} \left( c_{i, \min} + s_i \right)- R\right)  =& 0,\label{eqn:CS_1}\\[-4pt]
    \nu_i s_i =& 0,\label{eqn:CS_2}\\[-20pt]\nonumber
\end{align}
for all $i\in \mathcal{S}$. By solving (\ref{eqn:KKT_cond}) for $s_i$, we obtain
\begin{align}\nonumber\\[-20pt]\label{eqn:s_i_temp}
    s_i = C_i \left(\sqrt{\frac{A_i}{B_i(\theta- \nu_i)}} -1\right),\\[-20pt]\nonumber
\end{align}
where $A_i = \lambda_i (c_{i, \min} + \mu_i + \lambda_i)$, $B_i = c_{i,\min}\mu_i (\lambda_i+\mu_i)$, and $C_i = \frac{c_{i,\min}\mu_i}{\lambda_i+c_{i,\min}}$. By the C.S. condition (\ref{eqn:CS_2}), either $s_i > 0$ (which implies $\nu_i = 0$) or $s_i = 0$ (with $\nu_i \geq 0$). Thus, the optimal values of $s_i$, denoted by $s_i^*$, are equal to 
\begin{align}\nonumber\\[-20pt]\label{eqn:s_i_opt}
    s_i^* = C_i \left(\sqrt{\frac{A_i}{B_i\theta}} -1\right)^+,\\[-20pt]\nonumber
\end{align}
where $(x)^+ =  \max\{x,0\}$. 

Although the closed-form solution for each $s_i^*$ is given in \eqref{eqn:s_i_opt}, it depends on the Lagrange multiplier $\theta$, which must be chosen such that the conditional intensity budget constraint in \eqref{eqn:senders_problem_mod} is satisfied with equality, i.e., $\sum_{i\in\mathcal{S}}(c_{i,\min}+s_i^*)=R$. From \eqref{eqn:s_i_opt}, the total allocation $\sum_{i\in\mathcal{S}}(c_{i,\min}+s_i^*)$ is a continuous and strictly decreasing function of $\theta$ over $0<\theta\leq\max_{i\in\mathcal{S}}\frac{A_i}{B_i}$. As $\theta\to0^+$, the total allocation approaches infinity, whereas at $\theta=\max_{i\in\mathcal{S}}\frac{A_i}{B_i}$, it is equal to $\sum_{i\in\mathcal{S}}c_{i,\min}$. Therefore, if $\sum_{i\in\mathcal{S}}c_{i,\min}<R$, there exists a unique $\theta$ satisfying the conditional intensity budget, which can be obtained via bisection. At each iteration, we evaluate the total allocation at the midpoint of the current interval and update the bounds according to whether the resulting allocation is greater than or less than $R$. On the other hand, if $\sum_{i\in\mathcal{S}}c_{i,\min}>R$, the active set $\mathcal{S}$ is infeasible and can be discarded directly.

From the expression of $s_i^*$ in (\ref{eqn:s_i_opt}), the optimal
allocation for a given active set $\mathcal{S}$ exhibits a threshold
structure: only the sources with $\frac{A_i}{B_i} > \theta$ receive a
positive update rate. By Lemma~\ref{Lemma3}, a globally optimal policy
cannot allocate $c_i = c_{i,\min}$ and $s_i = 0$ to any source; hence, whenever the solution in (\ref{eqn:s_i_opt}) returns $s_i^* = 0$ for
some $i \in \mathcal{S}$, the candidate set $\mathcal{S}$ can be discarded
without loss of optimality, since by Lemma~\ref{Lemma3} no optimal solution
of (\ref{eqn:senders_problem_mod}) activates a source with a zero state-1
sampling rate. The sender's
globally optimal policy can therefore be found by enumerating the
candidate active sets, solving (\ref{eqn:s_i_opt}) for each, and eliminating the discarded sets.

\vspace{-0.15cm}  
\section{An Efficient Algorithm to Find the Sender's Optimal Information Revelation Policy}
\label{sec:efficient_algorithm}

In this section, we develop a branch-and-bound algorithm that finds the sender's globally optimal information-revelation policy with typically lower computational complexity than the exhaustive search described in Section~\ref{sect:multisourceproblem}. Recall from Lemma~\ref{Lemma_1} that, when $s_i>0$ and $c_i>0$, the sender's utility is strictly increasing in $s_i$ and strictly decreasing in $c_i$, whereas the PC requires the sender to allocate $c_i = c_{i,\min}$ to every source whose messages the receiver follows. Hence, the sender should ideally activate the sources that require a small $c_{i,\min}$ while providing a large utility in return. To identify such sources, we first express the sender's utility obtained from an active source $i$ (i.e., a source with $c_i = c_{i,\min}$ and $s_i>0$) in a more informative form.

Recall from (\ref{eqn:senders_problem_mod}) that the sender's utility obtained from source $i$ when $z_i = 1$ (i.e., $c_i = c_{i,\min}$) is given by
\begin{align}\label{eq:utility_active}
    J_{S,i}(s_i, c_{i,\min}) = \frac{\lambda_i}{\lambda_i + \mu_i}\, \frac{s_i (c_{i,\min} + \mu_i + \lambda_i)}{c_{i,\min}\mu_i + s_i(\lambda_i + c_{i,\min})}.
\end{align}
Since $c_i$ is fixed to $c_{i,\min}$ for all active sources, with a slight abuse of notation, in the remainder of this section we write $J_{S,i}(s_i)$ instead of $J_{S,i}(s_i, c_{i,\min})$. By using the constants $A_i$, $B_i$, and $C_i$ defined in (\ref{eqn:s_i_temp}), we can rewrite (\ref{eq:utility_active}) as
\begin{align}\nonumber\\[-19pt]
    J_{S,i}(s_i) = \frac{A_i C_i}{B_i}\, \frac{s_i}{C_i + s_i}. \label{eq:utility_three_param}\\[-19pt]\nonumber
\end{align}

We observe from (\ref{eq:utility_three_param}) that the utility contribution of an active source $i$ is fully characterized by three quantities: the marginal utility $\frac{A_i}{B_i}$, the minimum state-0 sampling rate $c_{i,\min}$, and the asymptotic ceiling of the sender's utility from source $i$ given by $\frac{A_i C_i}{B_i}$. The first quantity is the marginal utility $\frac{A_i}{B_i}$, which determines the initial rate of return at $s_i = 0$, i.e.,
\begin{align}\nonumber\\[-19pt]
    \frac{\partial J_{S,i}(s_i)}{\partial s_i}\bigg|_{s_i = 0} = \frac{A_i}{B_i} = \frac{\lambda_i}{(\lambda_i + \mu_i)\left(q\mu_i - (1-q)\lambda_i\right)}. \label{eq:marginal_utility}\\[-19pt]\nonumber
\end{align}
Since $J_{S,i}(s_i)$ is concave in $s_i$ by Lemma~\ref{Lemma4}, $\frac{A_i}{B_i}$ is also the largest marginal return that source $i$ can offer. Moreover, we recall from the KKT solution in (\ref{eqn:s_i_opt}) that, for a fixed active set, only the sources with $\frac{A_i}{B_i} > \theta$ receive a nonzero allocation $s_i^* > 0$. The second quantity is the minimum state-0 sampling rate $c_{i,\min} = \frac{q\mu_i}{1-q} - \lambda_i$, which is the fixed sampling rate that the sender must allocate to source $i$'s state-0 information in order to satisfy the PC. The third quantity is the asymptotic sender utility $\frac{A_i C_i}{B_i}$, which is the maximum utility that source $i$ can provide as $s_i \to \infty$, i.e.,
\begin{align}\nonumber\\[-19pt]
    \lim_{s_i \to \infty} J_{S,i}(s_i) = \frac{A_i C_i}{B_i} = \frac{\lambda_i}{q(\lambda_i + \mu_i)}. \label{eq:asymptotic_utility}\\[-19pt]\nonumber
\end{align}
Thus, while $\frac{A_i}{B_i}$ and $c_{i,\min}$ determine whether source $i$ receives a positive allocation, $\frac{A_i C_i}{B_i}$ limits how much utility source $i$ can ultimately contribute, even under an unlimited sampling rate. In the next lemma, we characterize how these three quantities vary with respect to $\lambda_i$, $\mu_i$, and $q$.

\begin{lemma}\label{lem:monotonicity}
    The marginal utility $\frac{A_i}{B_i}$ and the asymptotic utility $\frac{A_i C_i}{B_i}$ are both increasing in $\lambda_i$ and decreasing in $\mu_i$ and $q$, whereas the activation cost $c_{i,\min}$ is decreasing in $\lambda_i$ and increasing in $\mu_i$ and $q$.
\end{lemma}

\begin{Proof}
    We first show the monotonicity of $\frac{A_i}{B_i}$. The partial derivatives of $\frac{A_i}{B_i}$ with respect to $\lambda_i$ and $\mu_i$ are given by
    \begin{align*}\\[-19pt]
        \frac{\partial}{\partial \lambda_i}\left(\frac{A_i}{B_i}\right) &= \frac{q\mu_i^2 + (1-q)\lambda_i^2}{(\lambda_i + \mu_i)^2\left(q\mu_i - (1-q)\lambda_i\right)^2} > 0, \\[-3pt]
        \frac{\partial}{\partial \mu_i}\left(\frac{A_i}{B_i}\right) &= -\frac{\lambda_i\left((2q-1)\lambda_i + 2q\mu_i\right)}{(\lambda_i + \mu_i)^2\left(q\mu_i - (1-q)\lambda_i\right)^2} < 0,\\[-19pt]
    \end{align*}
    where the latter is strictly negative since $(2q-1)\lambda_i + 2q\mu_i = 2\left(q\mu_i - (1-q)\lambda_i\right) + \lambda_i > 0$ due to our initial assumption that $q\mu_i - (1-q)\lambda_i > 0$ (i.e., the receiver's default estimate is $\hat{x}_i(t) = 0$). In addition, we have
    \begin{align*}\\[-19pt]
        \frac{\partial}{\partial q}\left(\frac{A_i}{B_i}\right) = -\frac{\lambda_i}{\left(q\mu_i - (1-q)\lambda_i\right)^2} < 0,\\[-19pt]
    \end{align*}
    since $\lambda_i > 0$. Thus, $\frac{A_i}{B_i}$ is increasing in $\lambda_i$ and decreasing in $\mu_i$ and $q$. Next, for the asymptotic utility $\frac{A_i C_i}{B_i} = \frac{\lambda_i}{q(\lambda_i + \mu_i)}$, the partial derivatives are given by
    \begin{align*}\\[-20pt]
        \frac{\partial}{\partial \lambda_i}\left(\frac{A_i C_i}{B_i}\right) &= \frac{\mu_i}{q(\lambda_i + \mu_i)^2} > 0, \\
        \frac{\partial}{\partial \mu_i}\left(\frac{A_i C_i}{B_i}\right) &= -\frac{\lambda_i}{q(\lambda_i + \mu_i)^2} < 0, \\
        \frac{\partial}{\partial q}\left(\frac{A_i C_i}{B_i}\right) &= -\frac{\lambda_i}{q^2(\lambda_i + \mu_i)} < 0.\\[-20pt]
    \end{align*}
    Thus, the asymptotic utility $\frac{A_i C_i}{B_i}$ exhibits the same monotonic behavior as the marginal utility $\frac{A_i}{B_i}$.

    Finally, since $\frac{\partial c_{i,\min}}{\partial \lambda_i} = -1 < 0$, $\frac{\partial c_{i,\min}}{\partial \mu_i} = \frac{q}{1-q} > 0$, and $\frac{\partial c_{i,\min}}{\partial q} = \frac{\mu_i}{(1-q)^2} > 0$, the minimum state-0 sampling rate $c_{i,\min}$ is decreasing in $\lambda_i$ and increasing in $\mu_i$ and $q$, which completes the proof.
\end{Proof}

Lemma~\ref{lem:monotonicity} shows that a larger $\lambda_i$ and a smaller $\mu_i$ (i.e., a source that spends a larger fraction of time in state-1) simultaneously improve the initial slope $\frac{A_i}{B_i}$ and the asymptotic sender utility $\frac{A_i C_i}{B_i}$ while reducing $c_{i,\min}$. Similarly, a smaller $q$ makes persuasion easier in all three aspects: the marginal and asymptotic utilities increase while $c_{i,\min}$ decreases. This is consistent with the interpretation that a smaller $q$ means the receiver places less weight on correctly estimating state-0, which relaxes the PC and makes persuasion less costly for the sender. Based on these three quantities, we can now formally characterize when one source dominates another.

\begin{lemma}[Source Dominance] \label{lem:dominance}
    If source $i$ satisfies the following  conditions relative to source $j$, with at least one of them being strict
    \begin{align}\nonumber\\[-22pt]
        \frac{A_i}{B_i} &\geq \frac{A_j}{B_j}, \label{eq:dom_marginal} \\
        c_{i,\min} &\leq c_{j,\min}, \label{eq:dom_cost} \\
        \frac{A_i C_i}{B_i} &\geq \frac{A_j C_j}{B_j}, \label{eq:dom_ceiling}\\[-22pt]\nonumber
    \end{align}
    then no optimal solution to (\ref{eqn:senders_problem_mod}) can activate source $j$ while leaving source $i$ inactive. In this case, we say that source $i$ dominates source $j$ and express the dominance as $i \succ j$.
\end{lemma}

\begin{Proof}
    We prove the result in two steps. First, we show that conditions (\ref{eq:dom_marginal}) and (\ref{eq:dom_ceiling}) imply the point-wise utility dominance $J_{S,i}(s) \geq J_{S,j}(s)$ for all $s \geq 0$. Then, by using a swap argument together with condition (\ref{eq:dom_cost}), we prove the dominance claim.

    \emph{Step 1 (Point-wise utility dominance):} At $s = 0$, both utilities are equal to zero, and thus the inequality trivially holds. For any $s > 0$, from (\ref{eq:utility_three_param}), the condition $J_{S,i}(s) \geq J_{S,j}(s)$ is equivalent to
    \begin{align*}\\[-22pt]
        \frac{A_i C_i}{B_i}\, \frac{s}{C_i + s} \geq \frac{A_j C_j}{B_j}\, \frac{s}{C_j + s}.\\[-22pt]
    \end{align*}
    Since $s > 0$, $C_i + s > 0$, and $C_j + s > 0$, canceling $s$ and cross-multiplying yields the equivalent condition
    \begin{align*}\\[-22pt]
        \frac{A_i C_i}{B_i} (C_j + s) \geq \frac{A_j C_j}{B_j} (C_i + s).\\[-22pt]
    \end{align*}
    By noting that $C_k\! = \! \frac{A_k C_k}{B_k} \! \Big/ \! \frac{A_k}{B_k}$ for $k\!  \in\! \{i, j\}$ and rearranging the terms, the condition $J_{S,i}(s) \! \geq \! J_{S,j}(s)$ becomes 
    \begin{align}\nonumber\\[-22pt]
        \!\frac{A_i C_i}{B_i} \frac{A_j C_j}{B_j} \left(\!\frac{A_i}{B_i}\! -\! \frac{A_j}{B_j}\!\right)\!+\! \frac{A_i}{B_i} \frac{A_j}{B_j} s \!\left(\!\frac{A_i C_i}{B_i} \!-\! \frac{A_j C_j}{B_j}\!\right)\! \geq\! 0.\!\! \label{eq:dominance_decomp}\\[-22pt]\nonumber
    \end{align}
    The first term in (\ref{eq:dominance_decomp}) is non-negative since $\frac{A_i C_i}{B_i}, \frac{A_j C_j}{B_j} > 0$ and $\frac{A_i}{B_i} \geq \frac{A_j}{B_j}$ by (\ref{eq:dom_marginal}). Similarly, the second term is non-negative since $\frac{A_i}{B_i}, \frac{A_j}{B_j}, s > 0$ and $\frac{A_i C_i}{B_i} \geq \frac{A_j C_j}{B_j}$ by (\ref{eq:dom_ceiling}). Hence, we have $J_{S,i}(s) \geq J_{S,j}(s)$ for all $s \geq 0$. Furthermore, when at least one of the inequalities in (\ref{eq:dom_marginal}) or (\ref{eq:dom_ceiling}) is strict, the corresponding term in (\ref{eq:dominance_decomp}) is strictly positive, and thus $J_{S,i}(s) > J_{S,j}(s)$ for all $s > 0$.

    \emph{Step 2 (Swap argument):} Assume by contradiction that, in an optimal solution, source $j$ is active with $c_j = c_{j,\min}$ and $s_j > 0$ while source $i$ is inactive. Consider the following alternative policy: we deactivate source $j$, which frees a total sampling rate of $c_{j,\min} + s_j$, and activate source $i$ by allocating $c_i = c_{i,\min}$ and $s_i = s_j + (c_{j,\min} - c_{i,\min})$, while keeping the allocations of all the other sources unchanged. By condition (\ref{eq:dom_cost}), we have $c_{j,\min} \geq c_{i,\min}$, and thus $s_i \geq s_j > 0$. Moreover, the total sampling rate consumed by the new allocation is $c_{i,\min} + s_i = c_{j,\min} + s_j$, which is exactly equal to the freed rate; hence, the total sampling constraint remains satisfied.

    We now distinguish two cases depending on which of the conditions (\ref{eq:dom_marginal})--(\ref{eq:dom_ceiling}) holds with strict inequality. If at least one of (\ref{eq:dom_marginal}) or (\ref{eq:dom_ceiling}) is strict, then Step 1 gives $J_{S,i}(s_j) > J_{S,j}(s_j)$; since $J_{S,i}(\cdot)$ is strictly increasing by Lemma~\ref{Lemma_1} and $s_i \geq s_j$, we obtain $J_{S,i}(s_i) \geq J_{S,i}(s_j) > J_{S,j}(s_j)$. If instead (\ref{eq:dom_marginal}) and (\ref{eq:dom_ceiling}) hold with equality while (\ref{eq:dom_cost}) is strict, then (\ref{eq:dominance_decomp}) holds with equality, i.e., $J_{S,i}(s) = J_{S,j}(s)$ for all $s \geq 0$, but we have $s_i = s_j + (c_{j,\min} - c_{i,\min}) > s_j$; hence, $J_{S,i}(s_i) > J_{S,i}(s_j) = J_{S,j}(s_j)$ again by the strict monotonicity of $J_{S,i}(\cdot)$. In both cases, the alternative policy yields a strictly higher utility for the sender, contradicting the optimality of the original solution. Therefore, no optimal solution can activate source $j$ while leaving source $i$ inactive, which completes the proof.
\end{Proof}

The dominance relation $\succ$ defined through (\ref{eq:dom_marginal})--(\ref{eq:dom_ceiling}) forms a strict partial order among the sources, since it is irreflexive by the requirement of at least one strict inequality and transitive as the weak inequalities in (\ref{eq:dom_marginal})--(\ref{eq:dom_ceiling}) are preserved along a chain of comparisons. When the three conditions do not consistently favor one source over the other, the two sources are incomparable under $\succ$; in this case, their relative contributions cannot be determined a priori and must be evaluated directly through the objective function in (\ref{eqn:senders_problem_mod}).

Based on the dominance relation $\succ$ established in Lemma~\ref{lem:dominance}, we now develop a branch-and-bound algorithm that finds the optimal active set $\mathcal{S}^*$ without exhaustively enumerating all $2^n$ possible subsets. Throughout the search, the algorithm maintains three disjoint sets: $\mathcal{S}_{\text{in}}$, the set of sources decided to be active; $\mathcal{S}_{\text{out}}$, the set of sources decided to be inactive; and $\mathcal{S}_{\text{und}}$, the set of sources yet to be decided. At each step, the algorithm selects one source from $\mathcal{S}_{\text{und}}$ and branches into two subproblems: one in which the selected source is moved to $\mathcal{S}_{\text{in}}$ and one in which it is moved to $\mathcal{S}_{\text{out}}$. This branching induces a binary tree whose leaves correspond to fully determined active sets. To avoid exploring the entire tree, the algorithm employs three pruning mechanisms, which we describe next.

The first mechanism is \emph{dominance propagation}, which exploits the partial order $\succ$ obtained in Lemma~\ref{lem:dominance}. Whenever a source $j$ is moved to $\mathcal{S}_{\text{in}}$, every source $i \in \mathcal{S}_{\text{und}}$ with $i \succ j$ must also be moved to $\mathcal{S}_{\text{in}}$, since by Lemma~\ref{lem:dominance} no optimal solution can activate source $j$ while leaving source $i$ inactive. Conversely, whenever a source $i$ is moved to $\mathcal{S}_{\text{out}}$, every source $j \in \mathcal{S}_{\text{und}}$ with $i \succ j$ must also be moved to $\mathcal{S}_{\text{out}}$, since if the dominating source $i$ is not worth activating, then neither is the dominated source $j$. These forced decisions can cascade: a single inclusion or exclusion may trigger a chain of further forced decisions until no additional propagation is possible. After the propagation, if there exist $i \in \mathcal{S}_{\text{out}}$ and $j \in \mathcal{S}_{\text{in}}$ with $i \succ j$, then the current partial assignment contradicts Lemma~\ref{lem:dominance}, and the corresponding node is pruned.

The second mechanism is \emph{feasibility pruning}. If the total activation cost of the sources in $\mathcal{S}_{\text{in}}$ exceeds the conditional intensity budget, i.e., $\sum_{i \in \mathcal{S}_{\text{in}}} c_{i,\min} > R$, then no feasible completion of the current partial assignment exists, and the node is pruned.

The third mechanism is \emph{upper bound pruning}. At each node, we compute an upper bound, denoted by UB, on the utility achievable by any completion of the current partial assignment. If UB does not exceed the utility of the best solution found so far, then no completion can yield a higher utility, and the entire subtree rooted at the current node is pruned. To compute UB, we solve the KKT allocation in (\ref{eqn:s_i_opt}) over $\mathcal{S}_{\text{in}} \cup \mathcal{S}_{\text{und}}$, where the total rate available for the sampling rates $s_i$ is $R_{\text{rem}} = R - \sum_{i \in \mathcal{S}_{\text{in}}} c_{i,\min}$, that is, without charging the activation costs $c_{i,\min}$ of the undecided sources. This relaxation yields a valid upper bound as follows. Any feasible completion of the current node that activates a subset $\mathcal{T} \subseteq \mathcal{S}_{\text{und}}$ can allocate a total rate of at most $    R - \sum_{i \in \mathcal{S}_{\text{in}}} c_{i,\min} - \sum_{i \in \mathcal{T}} c_{i,\min} \leq R_{\text{rem}}$ to its sampling rates $s_i$, and only to the sources in $\mathcal{S}_{\text{in}} \cup \mathcal{T} \subseteq \mathcal{S}_{\text{in}} \cup \mathcal{S}_{\text{und}}$. Hence, the $s_i$ allocation of every feasible completion is also feasible for the relaxed problem, and consequently, UB is greater than or equal to the utility of the best achievable completion.

While the branching order does not affect the correctness of the algorithm, it influences its practical efficiency. As a heuristic, we branch on the sources in decreasing order of their marginal utility $\frac{A_i}{B_i}$. This choice tends to discover high-quality solutions early in the search, which in turn strengthens the upper bound pruning in the subsequent branches. The complete procedure is given in Algorithm~\ref{alg:bb}.

\begin{algorithm}[t]
\caption{Branch-and-Bound for the Optimal Active Set}
\label{alg:bb}
\begin{algorithmic}[1]
\Require $\{(\lambda_i, \mu_i)\}_{i=1}^{n}$, $q$, $R$
\Ensure Optimal active set $\mathcal{S}^*$ and utility $J^*$
\State Compute $\frac{A_i}{B_i}$, $c_{i,\min}$, $\frac{A_iC_i}{B_i}$ for all $i$; build the dominance relation $\succ$ via Lemma~\ref{lem:dominance}
\State $J^* \leftarrow 0$, $\mathcal{S}^* \leftarrow \emptyset$; \Call{Search}{$\emptyset$, $\emptyset$, $\{1, \ldots, n\}$}
\Procedure{Search}{$\mathcal{S}_{\text{in}},\, \mathcal{S}_{\text{out}},\, \mathcal{S}_{\text{und}}$}
\State Propagate $\succ$ until stable: if $j{\in}\mathcal{S}_{\text{in}}$, $i{\in}\mathcal{S}_{\text{und}}$, $i{\succ}j$, move $i{\to}\mathcal{S}_{\text{in}}$; if $i{\in}\mathcal{S}_{\text{out}}$, $j{\in}\mathcal{S}_{\text{und}}$, $i{\succ}j$, move $j{\to}\mathcal{S}_{\text{out}}$
\If{$\exists\, i \in \mathcal{S}_{\text{out}},\, j \in \mathcal{S}_{\text{in}}$ with $i \succ j$} \Return \EndIf
\If{$\sum_{i \in \mathcal{S}_{\text{in}}} c_{i,\min} > R$} \Return \EndIf
\If{$\mathcal{S}_{\text{in}} \neq \emptyset$ and $\sum_{i \in \mathcal{S}_{\text{in}}} c_{i,\min} < R$} solve (\ref{eqn:s_i_opt}) over $\mathcal{S}_{\text{in}}$ to obtain $J(\mathcal{S}_{\text{in}})$; \If{ $s_i^*>0$ for every $i\in \mathcal{S}_{\text{in}}$ and $J(\mathcal{S}_{\text{in}}) > J^*$}  $J^* {\leftarrow} J(\mathcal{S}_{\text{in}})$, $\mathcal{S}^* {\leftarrow} \mathcal{S}_{\text{in}}$;\EndIf \EndIf
\If{$\mathcal{S}_{\text{und}} = \emptyset$} \Return \EndIf
\State $R_{\text{rem}} \leftarrow R - \sum_{i \in \mathcal{S}_{\text{in}}} c_{i,\min}$; \textbf{if} $R_{\text{rem}} = 0$ \textbf{then} \Return
\State Solve (\ref{eqn:s_i_opt}) over $\mathcal{S}_{\text{in}} \cup \mathcal{S}_{\text{und}}$ with rate $R_{\text{rem}}$ for the $s_i$'s to obtain UB; \textbf{if} $\text{UB} \leq J^*$ \textbf{then} \Return
\State $k \leftarrow \argmax_{i \in \mathcal{S}_{\text{und}}} \frac{A_i}{B_i}$
\State \Call{Search}{$\mathcal{S}_{\text{in}} \cup \{k\}$, $\mathcal{S}_{\text{out}}$, $\mathcal{S}_{\text{und}} \setminus \{k\}$}
\State \Call{Search}{$\mathcal{S}_{\text{in}}$, $\mathcal{S}_{\text{out}} \cup \{k\}$, $\mathcal{S}_{\text{und}} \setminus \{k\}$}
\EndProcedure
\end{algorithmic}
\end{algorithm}
Since the pruning mechanisms only eliminate subtrees that provably cannot contain an optimal active set, Algorithm~\ref{alg:bb} is guaranteed to return the globally optimal solution of (\ref{eqn:senders_problem_mod}). Regarding the computational complexity, each node requires solving at most two KKT subproblems via the bisection search described in Section~\ref{sect:multisourceproblem}, each with a cost of $T_{\text{KKT}}(n)$, where $T_{\text{KKT}}(n)$ denotes the complexity of a single bisection-based KKT solution over at most $n$ sources. In the worst case, the algorithm visits $\mathcal{O}(2^n)$ nodes, resulting in a total runtime of $\mathcal{O}(2^n\, T_{\text{KKT}}(n))$. In this case, Algorithm~\ref{alg:bb} has the same complexity as the exhaustive search described in Section~\ref{sect:multisourceproblem}. In contrast, when the dominance relation $\succ$ forms a total order (i.e., a chain), the dominance propagation resolves all branching decisions, and the algorithm visits only $\mathcal{O}(n)$ nodes. For the intermediate cases, the number of visited nodes depends on the structure of the partial order, where the pruning mechanisms eliminate the subtrees corresponding to dominance-violating and suboptimal active sets.
\vspace{-0.20cm}
\section{The Optimal Information Revelation Policy for Multiple Receivers with Different Biases}
\label{sec:multi_receivers_bias}
\begin{figure}[tb]
  \centering
  \includegraphics[width=0.7\columnwidth]{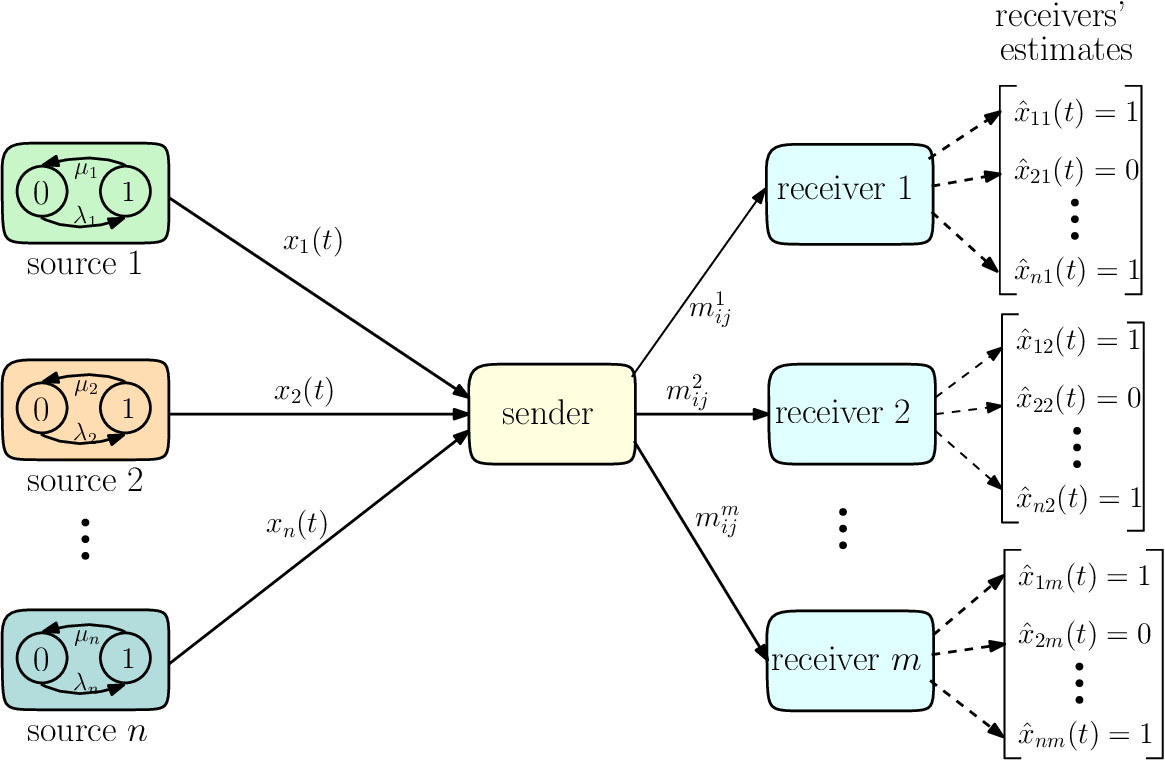}\vspace{-0.3cm}
  \caption{Communication system with $n$ sources, a sender, and $m$ receivers.}
  \label{fig:system2}
  \vspace{-0.35cm}
\end{figure}

\vspace{-0.10cm} 
In this section, we extend the single receiver model by considering $m\geq2$ receivers indexed by $r\in\{1,\dots,m\}$ and $n$ independent sources indexed by $i\in\{1,\dots,n\}$ as shown in Fig.~\ref{fig:system2}. Receiver $r$'s utility is characterized by substituting the parameter $q_r\in(0,1)$ in Table~\ref{Table:rec_utility}, which plays the same role as
$q$ in the single receiver formulation.   The sender observes the source(s) and commits to
Poisson sampling policies. If the sender satisfies the PC, each receiver $r$ forms an estimate about source $i$'s state denoted by $\hat{x}_{ir}(t)$ using the most recent message it receives, following the same procedure as in the single receiver model provided in (\ref{est_process}). If the sender fails to satisfy the PC for receiver $r$ regarding the state of source $i$, then receiver $r$ follows the default policy $\sigma_{\text{default}}$, under which it sets $\hat{x}_{ir}(t)=0$ for all $t$. 

We focus on the \emph{dedicated communication channel} setting in which the sender can communicate with each receiver over a private channel, and thus can select
receiver-specific sampling rates. Throughout this section, we use the same notation and utility definitions as in the single receiver
model, with the only modification that the receiver parameter is now indexed by $r$.
In particular, whenever the sender intends receiver $r$ to follow the information provided by
the sender (i.e., when a positive sampling rate for state-1 is used), receiver $r$'s
PC requirement induces a minimum state-0 sampling rate that depends on $q_r$ and source $i$'s parameters $\lambda_i$ and $\mu_i$. We assume that $q_r \mu_i > (1-q_r)\lambda_i$ for all $(i,r)$, so that each receiver's default estimate for every source is $\hat{x}_{ir}(t) = 0$, consistent with the single receiver assumption.

 Under \emph{dedicated communication channels} and the Poisson sampling policy introduced earlier, the sender can select a distinct Poisson sampling rate for each \emph{source–receiver pair} $(i,r)$, such that we have $(s_{ir},c_{ir})$ for $i=1,\dots,n$ and $r=1,\dots,m$. Here, $c_{ir}$ is the sampling rate when $x_i(t)=0$ and $s_{ir}$ is the sampling rate when $x_i(t)=1$.
The sender is subject to the conditional intensity budget given by $\sum_{i=1}^n\sum_{r=1}^m (s_{ir}+c_{ir})\le R$ with the feasibility constraints $s_{ir}\ge 0$ and $ c_{ir}\ge 0$. For each pair $(i,r)$, the same participation constraint argument as in the single receiver model applies, with
$q$ replaced by $q_r$. In particular, whenever we have $s_{ir}>0$,
it is necessary that $c_{ir}\ \ge\ c_{ir,\min}$ where $c_{ir,\min}\triangleq \frac{q_r\mu_i}{1-q_r}-\lambda_i.$ Since the sender utility for a fixed source is increasing in $s$ and decreasing in $c$ for $s>0$,
any feasible solution can be improved by setting $c_{ir}=c_{ir,\min}$ whenever $s_{ir}>0$.
Thus, the sender either keeps pair $(i,r)$ inactive by selecting $(s_{ir},c_{ir})=(0,0)$ or activates it with
$c_{ir}=c_{ir,\min}$ and $s_{ir}>0$. Under dedicated communication channels with multiple receivers, the sender maximizes the total utility across all pairs:
\begin{align}
\max_{\substack{\{\!s_{ir},c_{ir}\!\}}}
&\;\!\!\!\sum_{i=1}^n\!\sum_{r=1}^m \!J_{S,ir}(\!s_{ir},\!c_{ir}\!)\!\!=\!\! \sum_{i=1}^n\!\!\sum_{r=1}^m \!\!\frac{\lambda_i s_{ir}\,(c_{ir}+\lambda_i+\mu_i)}{(\!\lambda_i\!\!+\!\!\mu_i\!)\!\big(\!\mu_i c_{ir}\!\!+\!\!\lambda_i s_{ir}\!\!+\!\!c_{ir}s_{ir}\!\big)}\nonumber \\
\mbox{s.t.} \quad & \sum_{i=1}^n\sum_{r=1}^m (s_{ir}+c_{ir})\le R \nonumber \\
\quad &  c_{ir}\geq \mathbbm{1}(s_{ir}>0) c_{ir,\min}, \quad s_{ir} \geq 0\quad \forall i,r.
\label{eq:problem_private_multi_source_multi_recv}
\end{align}
Similar to Section~\ref{sect:multisourceproblem}, by introducing the binary activation variables $z_{ir}\!\!\in\!\!\{\!0,1\!\}$, we define the active set of pairs given by $\mathcal{S}\triangleq \{(i,r): z_{ir}=1\}.$ Then, if a source-receiver pair is selected to be active, we set $z_{ir}=1$, in which case the sender allocates $c_{ir}=c_{ir,\min}$ and $s_{ir}>0$. Otherwise, $z_{ir}=0$, and no resources are allocated, i.e., $c_{ir}=s_{ir}=0$. Following the approach in Section~\ref{sect:multisourceproblem}, we reformulate the optimization problem in (\ref{eq:problem_private_multi_source_multi_recv}) as
\begin{align}
\label{eq:problem_pairs_binary}
\max_{\{ \! z_{ir}, s_{ir} \! \}} \!\! & \hat{J}_S(\mathbf{z},\mathbf{s})  \!= \!\! \sum_{i=1}^{n} \!\sum_{r=1}^{m}  \!z_{ir}\frac{\lambda_i}{\lambda_i  \!\!+ \!\! \mu_i}  \frac{s_{ir} (c_{ir, \min} \!+\! \mu_i \!+\! \lambda_i)}{c_{ir, \min} \mu_i \!+\! s_{ir} \lambda_i \!+\! c_{ir, \min} s_{ir}} \nonumber \\
\mbox{s.t.} \quad & 
\sum_{i=1}^{n} \sum_{r=1}^m  \left( z_{ir} c_{ir, \min} + s_{ir} \right) \leq R\nonumber \\
\quad &  z_{ir} \in \{0, 1\}, ~s_{ir} \geq 0,\quad \forall i,r.
\end{align}
The optimization problem in \eqref{eq:problem_pairs_binary} has the same structure as the multi source  single receiver problem, with the difference that each decision variable is now over source-receiver pairs $(i,r)$. In particular, each pair $(i,r)$ is characterized by a fixed PC cost $c_{ir,\min}$ and a concave utility function $J_{S,ir}(s_{ir},c_{ir,\min})$ in terms of $s_{ir}$.
As in the previous analysis, for a fixed active set $\mathcal{S}$, the optimal allocation of
$\{s_{ir}\}_{(i,r)\in\mathcal{S}}$ admits a unique optimal solution. In particular, for each pair $(i,r)$, we 
define the constants which are obtained by substituting $c_{ir}=c_{ir,\min}$ into the single receiver
definitions and  $A_{ir} = \lambda_i (c_{ir,\min} + \mu_i + \lambda_i)$, $B_{ir} = c_{ir,\min}\mu_i (\lambda_i+\mu_i)$, and $C_{ir} = \frac{c_{ir,\min}\mu_i}{\lambda_i+c_{ir,\min}}$.
Then, for a fixed $\mathcal{S}$, there exists $\theta\ge 0$ such that
\begin{align}\nonumber\\[-21pt]
s_{ir}^*
=
C_{ir}\left(\sqrt{\frac{A_{ir}}{B_{ir}\,\theta}}-1\right)^+,
\qquad (i,r)\in\mathcal{S},
\label{eq:s_star_pairs}\\[-21pt]\nonumber
\end{align}
with $\theta$ selected so that $\sum_{(i,r)\in\mathcal{S}} s_{ir}^* = R-\sum_{(i,r)\in\mathcal{S}} c_{ir,\min}$.

Consequently, each pair $(i,r)$ can be expressed by the same three quantities that govern the branch-and-bound procedure in Algorithm~\ref{alg:bb}. The first quantity is the marginal utility $A_{ir}/B_{ir}$, the second is the minimum state-0 sampling rate $c_{ir,\min}$, and the third is the asymptotic sender utility $A_{ir}C_{ir}/B_{ir}$ from source $i$ and receiver $r$, all of which depend on both the receiver bias $q_r$ and the source parameters $(\lambda_i,\mu_i)$. The dominance relation in Lemma~\ref{lem:dominance} applies directly after replacing the source index $i$ by the pair index $(i,r)$: pair $(i_1, r_1)$ dominates pair $(i_2, r_2)$ if the three conditions~\eqref{eq:dom_marginal}--\eqref{eq:dom_ceiling} hold with at least one strict inequality when evaluated at the pairwise quantities.
 
The optimization problem in~\eqref{eq:problem_pairs_binary} can be viewed as a multi source-receiver pair selection problem with $N = nm$ items, indexed by $(i,r)$. For each source-receiver pair, the utility function in~\eqref{eq:problem_pairs_binary} is of the same form as in the multi source  single receiver formulation. Algorithm~\ref{alg:bb} is then applied by treating each pair $(i,r)$ as an individual item, with the same dominance propagation, feasibility pruning, and upper bound pruning operating over the $N$ pairs. Thus, Algorithm~\ref{alg:bb} has the following search complexity: $i$) in the worst case when there is no dominance relation between the source-receiver pairs, its search complexity is $\mathcal{O}(2^{N} T_{\text{KKT}}(N))$; and $ii$) in the best case when the $(i,r)$ pairs admit a perfect ordering, the search complexity reduces to $\mathcal{O}(N T_{\text{KKT}}(N))$.  The corresponding objective values are computed by solving the inner KKT allocation in~\eqref{eq:s_star_pairs} for each candidate active set under consideration. 

\vspace{-0.25cm}
\section{Numerical Results}\label{Num_result}
\vspace{-0.15cm}
\begin{figure}[t]
  \centering

  \includegraphics[width=.79\columnwidth]{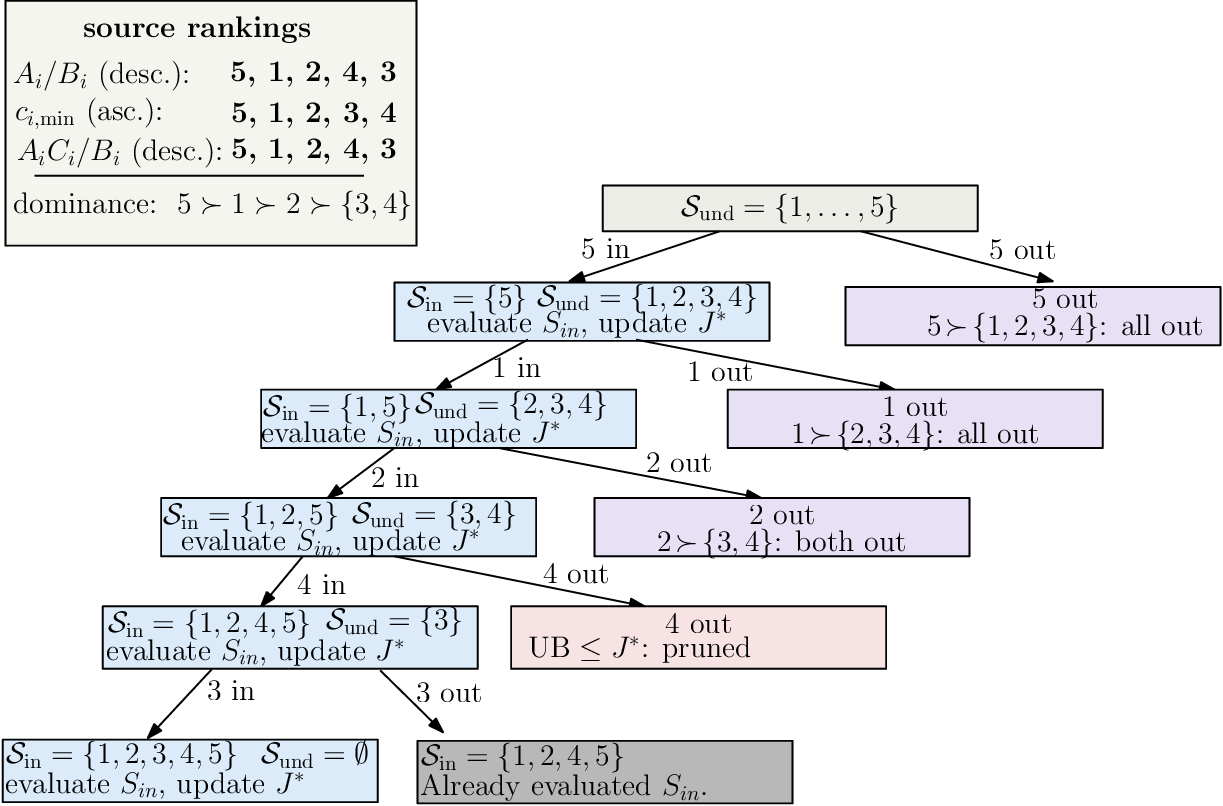}
   \vspace{-0.35cm}
  \caption{Branch-and-bound search for the five-source example.}
  \label{fig:algoexample}
  \vspace{-0.35cm}
\end{figure}
In this section, we provide numerical results to illustrate the theoretical results and the behavior of the proposed allocation method.
\vspace{-0.35cm}
\subsection{Multi source and Single receiver Case}\label{subsec:num_res_1}
\vspace{-0.15cm}
In our first numerical example, we consider \(n=5\) sources with transition rates
\(\lambda=[1.3,\,1.8,\,0.7,\,2.3,\,1.5]\),
\(\mu=[2.3,\,3.8,\,3.2,\,5.3,\,2.0]\), and receiver bias \(q=0.5\).
For these parameters, we compute the three source parameters introduced in Section~\ref{sec:efficient_algorithm}, namely the marginal utility \(A_i/B_i\), the minimum state-0 sampling rate \(c_{i,\min}\), and the asymptotic sender utility \(A_iC_i/B_i\) obtained from source $i$. The corresponding rankings are shown in Fig.~\ref{fig:algoexample}. In particular, the marginal-utility ranking and the asymptotic-utility ranking both yield the order \(5,1,2,4,3\), whereas the minimum state-0 sampling $c_{i,\min}$ rate ranking yields \(5,1,2,3,4\). Hence, sources \(5\), \(1\), and \(2\) admit a consistent dominance ordering, while sources \(3\) and \(4\) are incomparable under Lemma~\ref{lem:dominance}. The resulting dominance structure is therefore \(5 \succ 1 \succ 2\), with both sources \(3\) and \(4\) dominated by source \(2\) but incomparable with each other.

This dominance structure substantially reduces the effective search space. Indeed, among the \(2^5-1=31\) nonempty active sets, only the following six sets are dominance-closed:
\(\{5\}\), \(\{1,5\}\), \(\{1,2,5\}\), \(\{1,2,3,5\}\), \(\{1,2,4,5\}\), and \(\{1,2,3,4,5\}\).
The branch-and-bound search tree generated by Algorithm~\ref{alg:bb} is shown in Fig.~\ref{fig:algoexample}. The algorithm branches according to decreasing marginal utility \(A_i/B_i\), starting from source \(5\). The leftmost explored (blue) branch corresponds to the successive inclusion of sources \(5\), \(1\), \(2\), \(4\), and \(3\). At each visited blue node, the current active set \(\mathcal{S}_{\mathrm{in}}\) is evaluated and the best known value \(J^*\) is updated whenever an improvement is obtained.

The main computational savings arise from dominance propagation. Whenever a source is excluded, all sources dominated by it are also excluded, and the corresponding sub-tree (purple boxes) is collapsed without further evaluation. For example, excluding source \(5\) at the root forces the exclusion of all remaining sources, since \(5 \succ 1 \succ 2\) and \(2 \succ \{3,4\}\). Similarly, excluding source \(1\) forces the exclusion of sources \(2\), \(3\), and \(4\), while excluding source \(2\) forces the exclusion of sources \(3\) and \(4\). These cascaded eliminations account for most of the pruning observed in Fig.~\ref{fig:algoexample}. Upper-bound pruning (red boxes) provides an additional reduction when dominance alone is insufficient. In particular, after reaching the node with \(\mathcal{S}_{\mathrm{in}}=\{1,2,5\}\) and exploring the branch in which source \(4\) is excluded, the algorithm computes an optimistic upper bound by solving the relaxed KKT allocation over \(\{1,2,3,5\}\), i.e., without considering $c_{3,\min}$ of the undecided source \(3\). Since this upper bound does not exceed the best known value already attained at \(\mathcal{S}_{\mathrm{in}}=\{1,2,4,5\}\), that sub-tree is pruned.

The node corresponding to the active set \(\{1,2,4,5\}\) in Fig.~\ref{fig:algoexample} is the optimal solution for this example which highlights the importance of the asymptotic utility \(A_iC_i/B_i\). Although source \(4\) has a higher minimum state-0 sampling rate than source \(3\), namely \(c_{4,\min}=3.0\) versus \(c_{3,\min}=2.5\), it is preferred in the optimal active set because its asymptotic utility is substantially larger, i.e., \(A_4C_4/B_4=0.605\) versus \(A_3C_3/B_3=0.359\).
Overall, the branch-and-bound algorithm visits only 11 search-tree nodes for this example, whereas exhaustive search would require evaluating all 31 nonempty active sets.

\begin{figure}[t]
  \centering
  \includegraphics[width=.63\columnwidth]{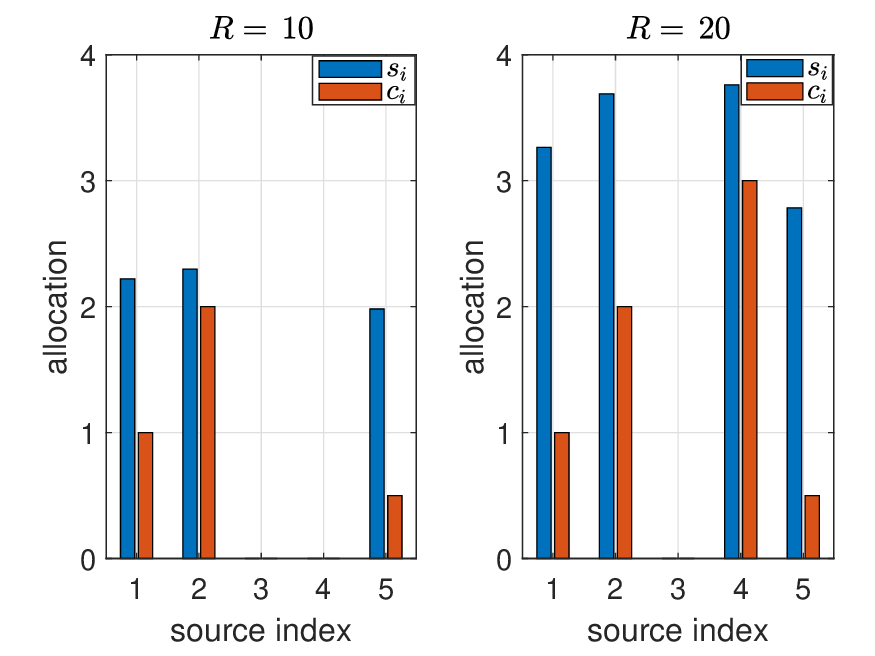}\vspace{-0.35cm}
  \caption{The optimal resource allocation of the sender for $R=\{10,20\}$.}\label{fig:AllocationFigure}
  \vspace{-0.4cm}
\end{figure}

With the same \((\lambda_i,\mu_i)\) sets given previously and $q=0.5$, we plot the optimal
\((s_i,c_i)\) allocations in Fig.~\ref{fig:AllocationFigure} when the conditional intensity budget is $R=\{10,20\}$. For \(R=10\), the optimal active set is
\(\{1,2,5\}\), i.e., the three most dominant sources shown in Fig.~\ref{fig:algoexample}, and the corresponding bars show
positive \(s_i\) only on these indices with \(c_i=c_{i,\min}\). When we increase
the budget to \(R=20\), the sender starts to send updates about source \(4\)'s state and the optimal set becomes
\(\{1,2,4,5\}\). This example also illustrates the necessity of our branch-and-bound algorithm. If sources were ranked solely based on their $c_{i,\min}$ values, source $4$, which has a relatively large $c_{i,\min}$, would appear to be the least efficient source. We would therefore consider $\{1,2,3,5\}$ followed by $\{1,2,3,4,5\}$, thereby missing the optimal set for $R=20$, namely $\{1,2,4,5\}$.
\begin{figure}[t]
  \centering
\includegraphics[width=.70\columnwidth]{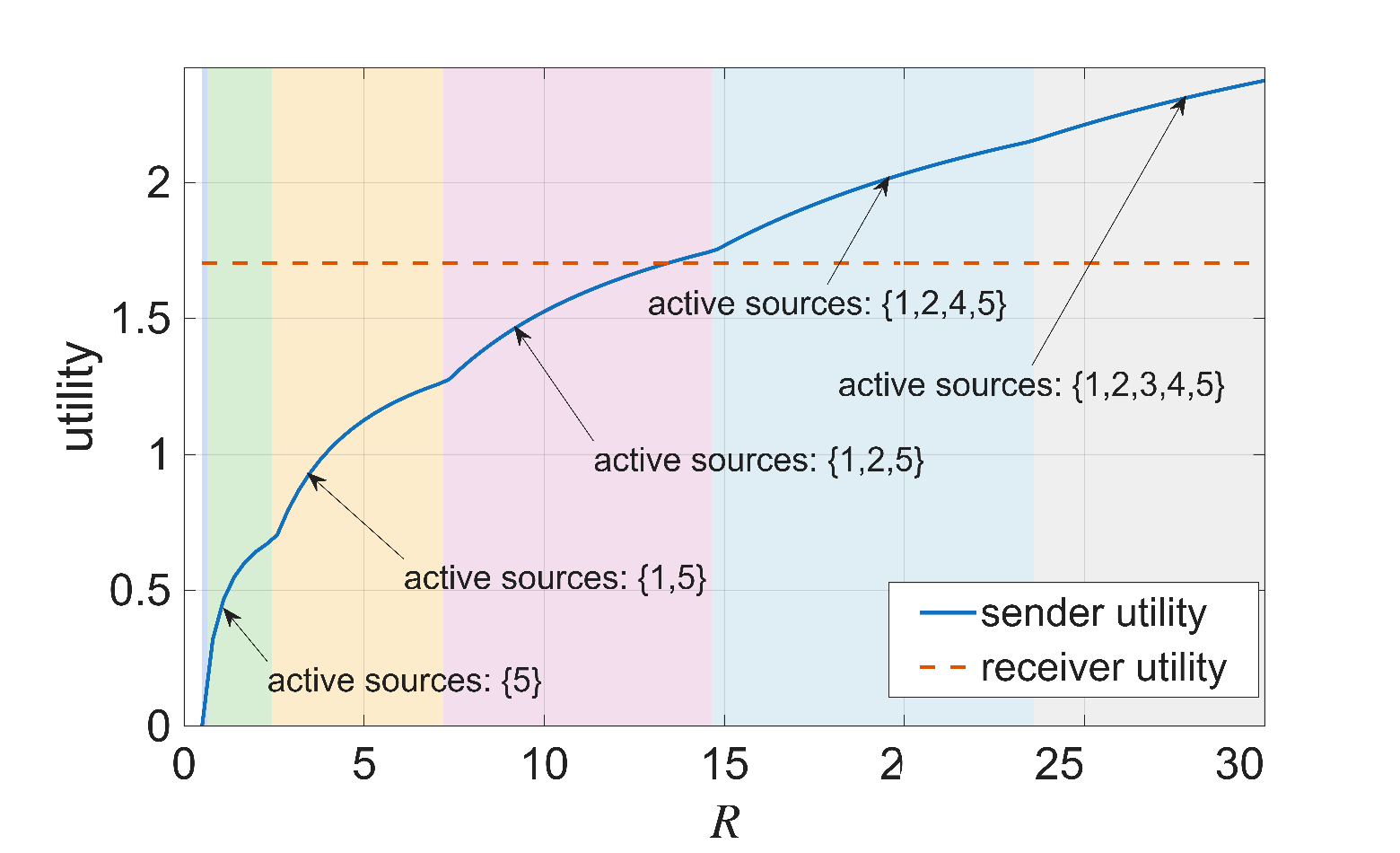}\vspace{-0.35cm}
  \caption{Utility functions  with respect to the conditional intensity budget $R$. } 
  \label{fig:utilityvsR}
  \vspace{-0.4cm}
\end{figure}

Next, we plot the sender's optimal utility \(J_S^*\) as the budget \(R\) increases, together with the receiver's benchmark utility, which remains constant, as shown in Fig.~\ref{fig:utilityvsR}. We consider the same set of $\lambda_i$ and $\mu_i$ as before and choose $q=0.5.$ 
The colored bands highlight the different regions which are the intervals of \(R\)  where the
optimal active source set remains the same. For example, when the conditional intensity budget is very limited, i.e., $0.5< R\leq {2.55}$, the sender can only send updates about the most dominant source (source 5). As 
$R$ increases, the sender begins to transmit updates about a larger set of sources, specifically in the following order: 
\(\{5\}\), \(\{1,5\}\), \(\{1,2,5\}\), \(\{1,2,4,5\}\), and finally
\(\{1,2,3,4,5\}\). Each transition occurs exactly when allocating \(c_{i,\min}\) to the next most dominant source, along with some $s_i$ on that source, becomes more beneficial than further increasing the $s_i$ of currently active sources. The receiver's utility curve remains constant by the construction of the problem due to the optimistic leader-favorable tie-breaking in (\ref{eqn:Stackelberg_2}). As we increase the sender's conditional intensity budget $R$, the sender can obtain a higher utility whereas the receiver's
default utility remains constant across all \(R\).

\begin{figure}[t]
  \centering  \includegraphics[width=.62\columnwidth]{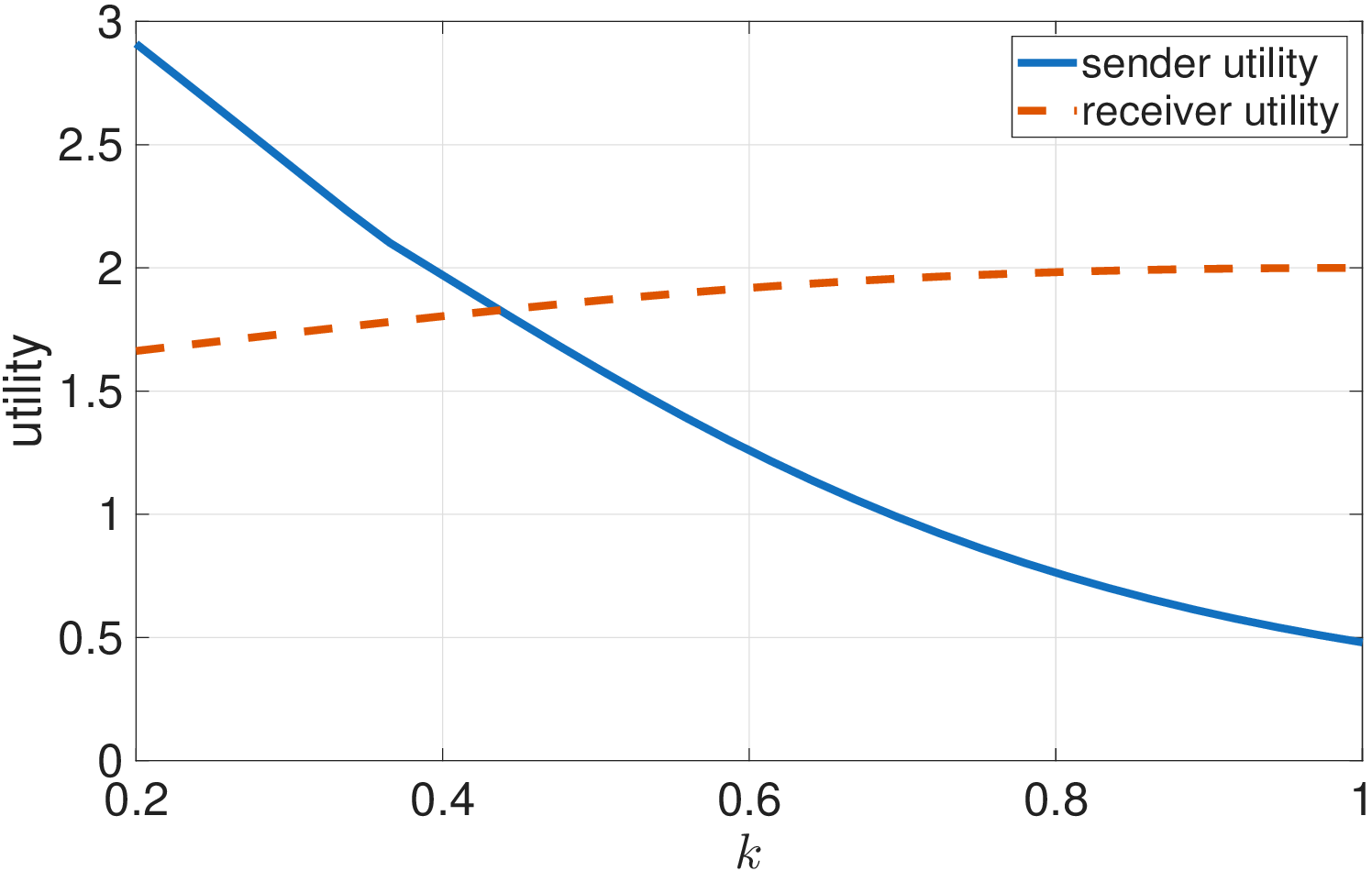}
  \vspace{-0.3cm}
  \caption{Agents' utilities with respect to heterogeneity parameter of $k$. }\label{fig:heterok}\vspace{-0.35cm}
\end{figure}
In our next simulation result, we choose \(n=5\), \(q=0.5\), \(R=15\), and \(\lambda_i= 1\) for all $i$.
To ensure the PC, we impose \(\mu_i\ge 1\).  We generate heterogeneous
\(\mu_i\)-profiles by
\begin{align*}\\[-21pt]
   \mu_i \;=\; 1+\,(C-n)\,\frac{k^i}{\sum_{j=1}^n k^j},\\[-21pt] 
\end{align*}
so that \(\sum_i \mu_i=C=20\) and each \(\mu_i\ge 1\). We vary the parameter \(k\in[0.2,1]\) to control heterogeneity of the $\mu_i$ distribution. For example, \(k=1\) yields the uniform case where all \(\mu_i\)'s are equal, while
smaller \(k\) values produce a more uneven distribution, with some \(\mu_i\) values close to 1 and others significantly larger.
The blue solid curve in Fig.~\ref{fig:heterok} shows the optimal sender utility
\(J_S^*\) as a function of \(k\) at the equilibrium.  The minimum sender utility occurs at \(k=1\)
(where \(\mu_i = 4\) for all $i$), and the sender's utility increases as \(k\) decreases (which leads to a more
heterogeneous distribution of $\mu_i$).  Intuitively, with \(q=0.5\) and \(\lambda_i=1\), the PC for source \(i\) is \(c_{i,\min}=\mu_i-1\).  A more heterogeneous distribution of $\mu_i$ creates some
sources with \(\mu_i\) close to \(1\), hence leads to very small \(c_{i, \min}\) values for some sources. These sources are easier to activate as they do not require high \(c_{i, \min}\) and also yield high marginal utility for persuasion. Thus, the sender benefits from the heterogeneous distribution of $\mu_i$'s. The red dashed curve in Fig.~\ref{fig:heterok} shows the receiver's utility
\(\sum_i \tfrac{q\,\mu_i}{\mu_i+\lambda_i}\). Because \(f(\mu_i)=\mu_i/(\mu_i+1)\) is
\emph{concave} in \(\mu_i\),  the receiver's utility (\(q\sum_i  f(\mu_i)\)) is maximized with the uniform $\mu_i$ distribution and
decreases with heterogeneity in $\mu_i$. Thus, the heterogeneity in $\mu_i$ helps the sender (by lowering some \(c_{i,\min}\)) but hurts the receiver, yielding the
opposing trends shown in Fig.~\ref{fig:heterok}.
\vspace{-0.35cm}
\subsection{Multi source and Multi receiver Case}
\vspace{-0.15cm}

We consider the multi source  multi receiver setting and illustrate the optimal pairwise utility contributions under different total budgets. We use \(4\) sources with parameters
\(\lambda=[0.25,\,0.4,\,0.8,\,1.1]\) and
\(\mu=[1.1,\,1.4,\,1.5,\,1.82]\), and \(4\) receivers with utility weights
\(q_r=[0.47,\,0.38,\,0.62,\,0.55]\). Here, we solve the joint source--receiver allocation problem using Algorithm~\ref{alg:bb} and compute the optimal per-pair utility values under the conditional intensity budget $\!R$.

\begin{figure}[t]
  \centering
  \subfloat[]{\includegraphics[width=0.32\columnwidth]{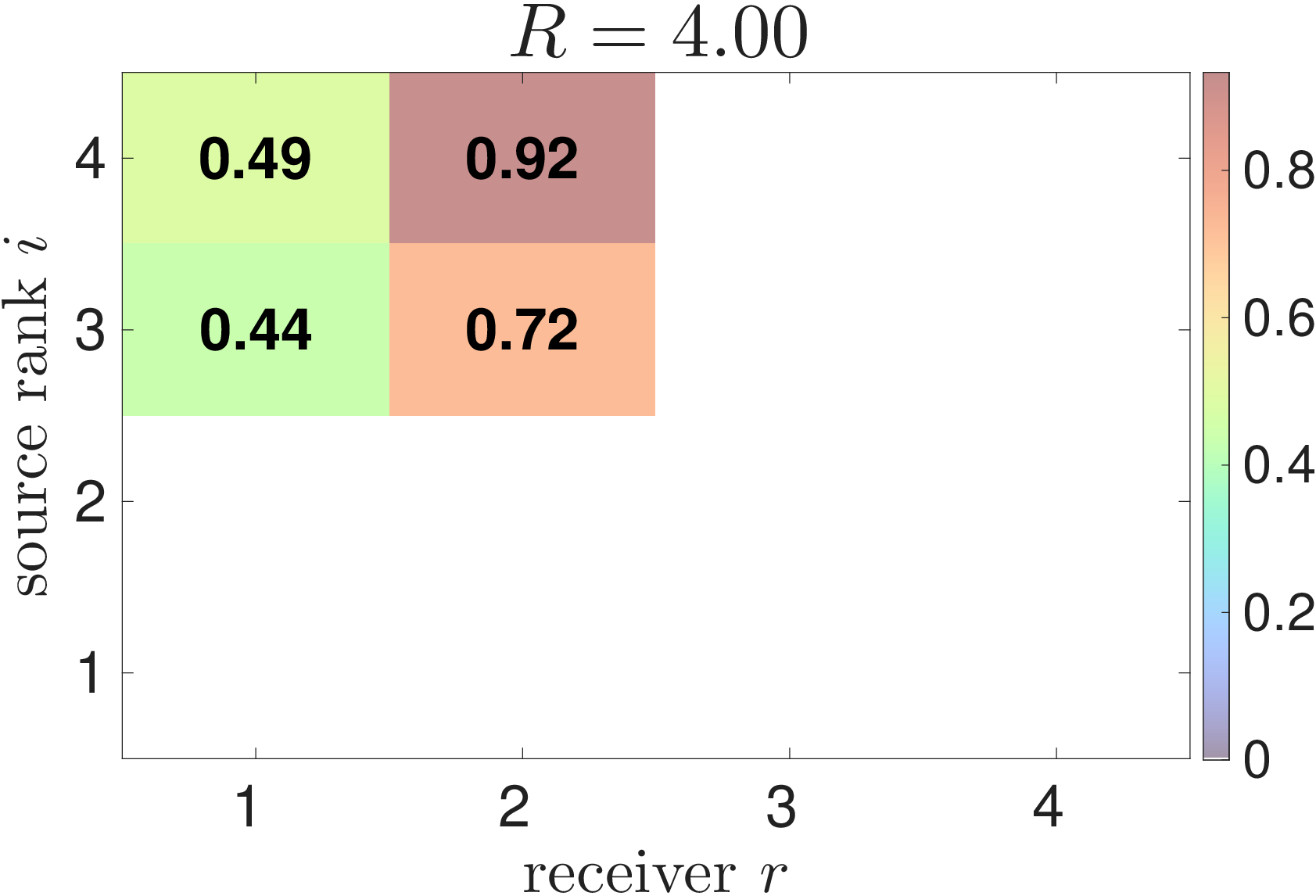}\label{fig:numres6a}}~
  \subfloat[]{\includegraphics[width=0.32\columnwidth]{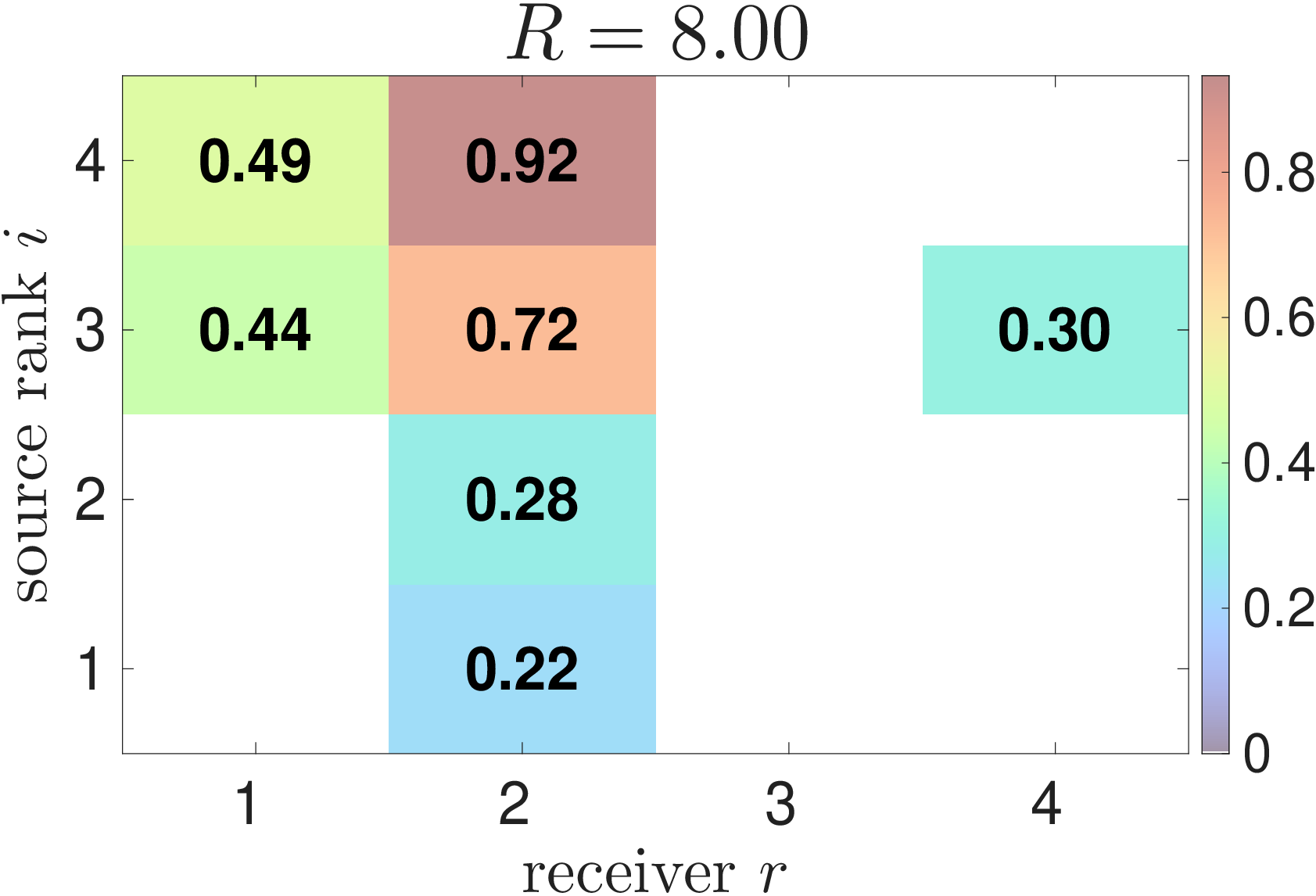}\label{fig:numres6b}}~
  \subfloat[]{\includegraphics[width=0.32\columnwidth]{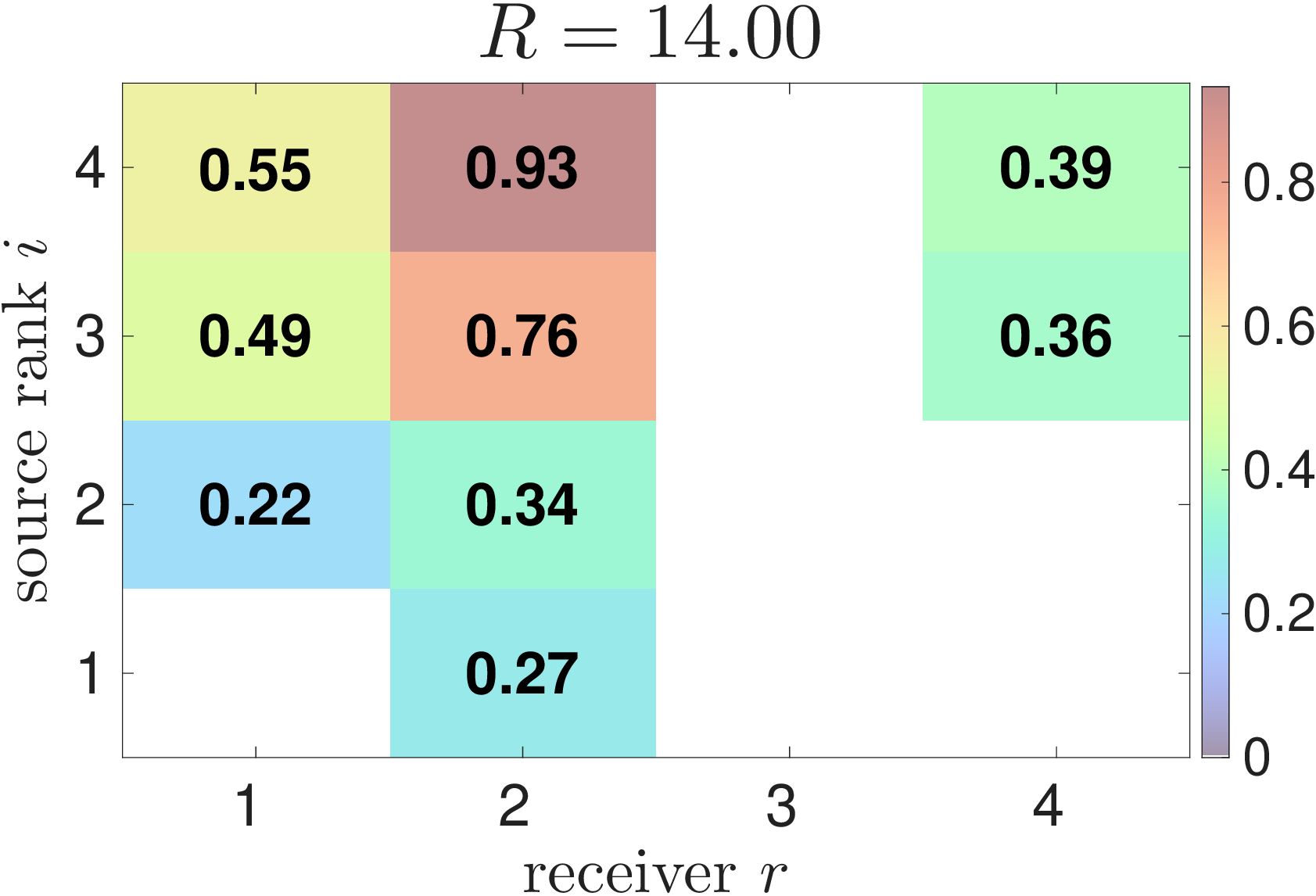}\label{fig:numres6c}}
\vspace{-0.26cm}
  \caption{Multi source multi receiver: per-pair utility heatmaps under different budgets.}
  \label{fig:numres6}
  \vspace{-0.4cm}
  \end{figure}
  Fig.~\ref{fig:numres6} shows the resulting heatmaps of per-pair utility contributions for  \(R=\{4,8,14\}\). Each nonzero cell in the heatmap corresponds to an active source-receiver pair, and the number written in the cell is the utility contribution of that pair to the sender's total utility. In the low-budget regime (when \(R=4\)), as shown in Fig.~\ref{fig:numres6}(a), the optimal policy is highly selective and activates only a small number of source-receiver pairs. The allocation is concentrated on a few dominant pairs such as $(i,r)= (4,2)$ and $(3,2)$. The corresponding utility values are relatively large compared to the number of active links. This behavior is consistent with the insight from the single receiver case: when the total budget is limited, the sender prioritizes only the most dominant source-receiver pairs and leaves the remaining pairs inactive. When the budget increases to \(R=8\) (shown in Fig.~\ref{fig:numres6}(b)), additional source-receiver pairs become active. In this example, the sender first retains the previously selected pairs with relatively high utility contributions and then expands the active set by including additional pairs with smaller but still positive contributions. For the larger budget \(R=14\) (as shown in Fig.~\ref{fig:numres6}(c)), some source-receiver pairs continue to have higher utility contributions than others while increasing their total contributed utility compared to the lower budget cases. The persistence of several high-valued cells across all three panels shows that certain source-receiver pairs, such as $(4,2)$ and $(3,2)$, receive larger utility allocations for the chosen  $(\lambda_i,\mu_i,q_r)$ values.
\begin{figure}[t]
  \centering  \includegraphics[width=.70\columnwidth]{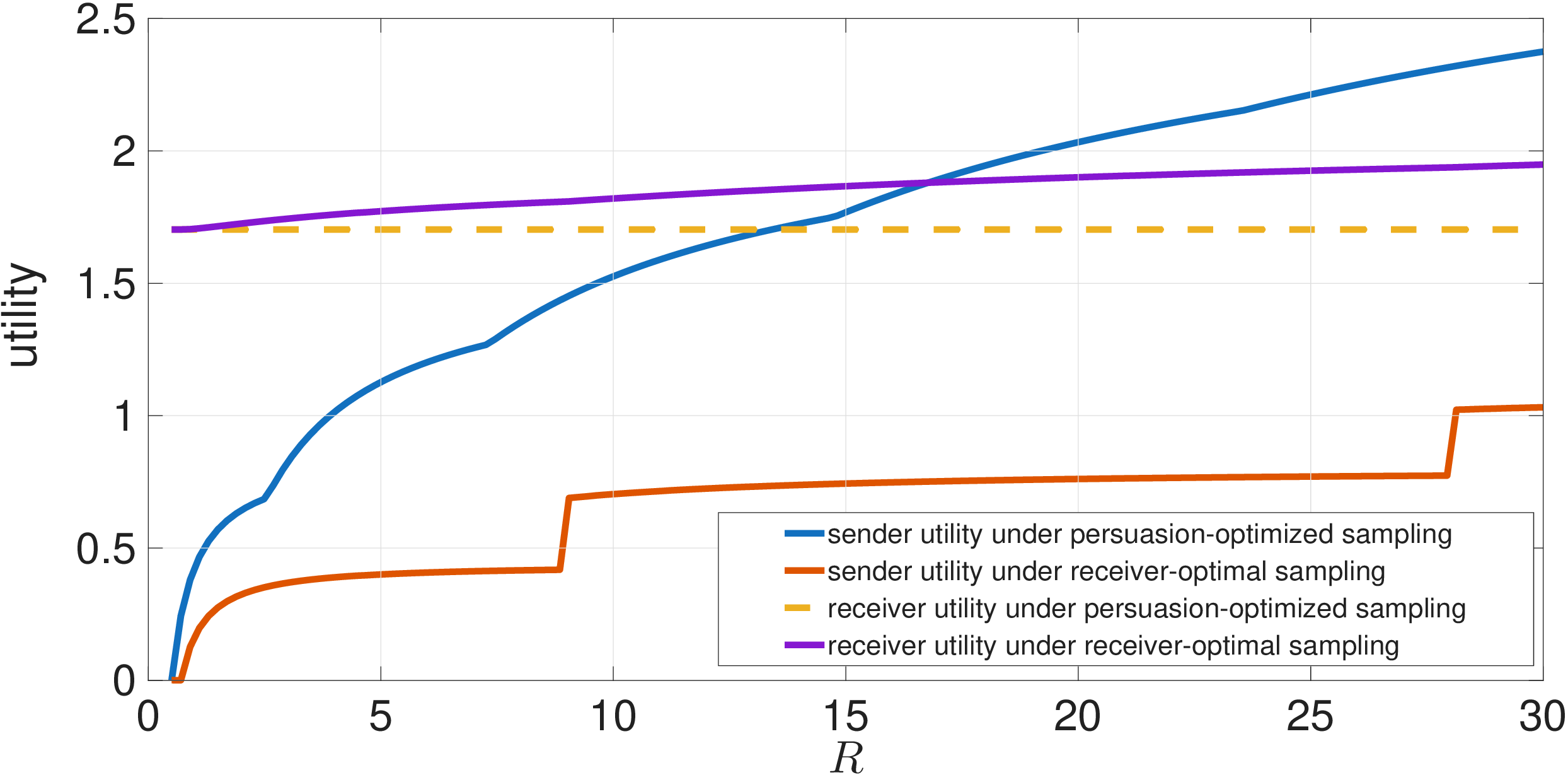}
  \vspace{-0.2cm}
  \caption{Utility comparison under sender-optimal persuasion and receiver-optimal sampling.}
  \label{fig:numres9}
  \vspace{-0.3cm}
\end{figure}
\vspace{-0.25cm}
\subsection{\!\!Persuasion-Optimized versus Receiver-Optimized Sampling }\label{subsec:disc_persuasion_vs_nopersuasion}

In Fig.~\ref{fig:numres9}, we compare the optimal utilities under our persuasion model with a receiver-optimized sampling benchmark in which the sender directly maximizes the receiver's weighted steady-state correct estimation probability by solving problem (\ref{eqn:nopersuasion_problem}) via exhaustive search for the same \(n=5\) sources and \((\lambda_i,\mu_i)\) parameters as in Subsection~\ref{subsec:num_res_1}, with receiver bias \(q=0.5\). For the receiver-optimized benchmark, we define the extended
per-source receiver utility as
\begin{align*}
\bar{J}_{R,i}(s_i,c_i)
\triangleq
\begin{cases}
\displaystyle
\frac{
q\mu_i c_i(\mu_i+s_i)
+(1-q)\lambda_i s_i(\lambda_i+c_i)
}{
(\mu_i+\lambda_i)
\bigl(\mu_i c_i+\lambda_i s_i+c_i s_i\bigr)
},
\\\hspace{3.7cm} \text{ if }(s_i,c_i)\neq(0,0),\\
\displaystyle
\frac{q\mu_i}{\mu_i+\lambda_i},
 \hspace{2.4cm} \text{ if }(s_i,c_i)=(0,0).
\end{cases}
\end{align*}
For an inactive source with $(s_i,c_i)=(0,0)$, no updates are
transmitted, and the receiver therefore employs the prior-only
default estimator $\sigma_{\mathrm{default}}$. Accordingly, its
utility is defined as the prior utility
$q\mu_i/(\mu_i+\lambda_i)$. Then, we write the receiver-optimized sampling rate allocation problem as:
\begin{align}\nonumber\\[-20pt]
\label{eqn:nopersuasion_problem}
\!\max_{\{\! s_i, c_i\! \}}  \quad &
\sum_{i=1}^{n}\bar{J}_{R,i}(s_i,c_i)
\nonumber\\[-3pt]
\mbox{s.t.} \quad &
\sum_{i=1}^{n} (c_i + s_i) \leq R \nonumber\\[-3pt]
\quad & c_i \geq 0,\quad s_i \geq 0,\qquad i\in\{1,\ldots,n\}. \!\!\!\\[-20pt]\nonumber
\end{align}
Fig.~\ref{fig:numres9} shows that the receiver utility under the receiver-optimal sampling benchmark (purple dash-dotted curve) is higher (in the plotted regime) than the receiver utility in the persuasion setup (yellow dashed curve), since without persuasion the budget can be used entirely to optimize the correct state estimation accuracy. However, in the persuasion setup the sender chooses \(c_i=c_{i,\min}\) for active sources and, under the optimistic assumption we have adopted in this problem, the sender keeps the receiver exactly at its baseline utility while using the remaining budget to maximize its own objective through the \(s_i\) allocations. Turning to the sender utilities, the persuasion model (blue solid curve) exhibits the active-set structure identified earlier: the utility increases with \(R\), with visible breakpoints at each change of the optimal active set. The red solid curve shows the sender utility in (\ref{eqn:senders_problem}) evaluated at the receiver-optimal allocation, that is, the \((s_i, c_i)\) obtained by solving problem (\ref{eqn:nopersuasion_problem}), when the receiver employs $\sigma_{\text{sender}}$. This curve lies strictly below the persuasion-model utility, since the receiver-optimal allocation is not designed to favor the sender; moreover, its visible jumps occur when the receiver-optimal exhaustive solution activates a new source with nontrivial \((s_i,c_i)\), which creates a discrete improvement in sender utility, while the receiver-optimal curve itself remains comparatively smooth since it is the quantity being directly optimized. Overall, Fig.~\ref{fig:numres9} shows the main strategic benefit of the timely persuasion: by paying only the minimum communication cost needed to secure receiver compliance, the sender can use more of the budget for state-1 sampling and thereby achieve substantially higher sender utility, even though this keeps the receiver at its minimum acceptable base utility level.
\vspace{-0.45cm}
\section{Conclusion}
\vspace{-0.20cm}

In this paper, we studied a dynamic persuasion problem in which a sender influences a receiver by controlling the timing of information updates from binary CTMC sources. We derived closed-form participation constraints and obtained an explicit solution for the single source case. For the multi source  setting under a conditional intensity budget, we developed a bisection procedure for the
unique KKT multiplier to compute the optimal state-dependent update rates for a fixed active set, and designed an efficient algorithm to determine the optimal active sources with potentially lower search complexity. We further extended the model to a multi receiver setting with heterogeneous receiver biases and solved the problem using the same algorithmic framework. Numerical results illustrated how the sender allocates the budget across source--receiver pairs, how active sets evolve with the conditional intensity budget, and how persuasion increases the sender’s utility compared to the receiver-optimal sampling benchmark. As a future direction, we plan to extend the model to multi sender environments and study equilibrium policies under both aligned and misaligned sender objectives.

\vspace{-0.45cm}
\bibliographystyle{IEEEtran}
\bibliography{IEEEabrv,refs}

\end{document}